\documentclass[twocolumn]{aastex61}
\usepackage{amsmath}
\usepackage{longtable}
\usepackage{threeparttablex}
\usepackage{bm}

\renewcommand{\sectionautorefname}{\S}

\newcommand{\ms}{\ensuremath{\rm m\,s^{-1}}}

\newcommand{\rhk}{\ensuremath{R^{\prime}_{\rm HK}}}	
\newcommand{\logrhk}{\ensuremath{\log\rhk}}		
\newcommand{\shk}{\ensuremath{S_{\rm HK}}}	

\newcommand{\mjup}{\ensuremath{M_{\rm J}}}

\newcommand{\thisstar}{$\epsilon$ Eridani}
\newcommand{\radvel}{\texttt{RadVel}}
\newcommand{\emcee}{\texttt{emcee}}
\newcommand{\VIP}{\texttt{VIP}}
\newcommand{\PyKLIP}{\texttt{PyKLIP}}

\received{May 1, 2018}
\revised{July 15, 2018}
\accepted{\today}

\submitjournal{AJ}
\shorttitle{Deep exploration of $\epsilon$ Eridani}
\shortauthors{Mawet et al.}

\begin{document}

\title{Deep exploration of $\epsilon$ Eridani with Keck Ms-band vortex coronagraphy and radial velocities: mass and orbital parameters of the giant exoplanet\footnote{Based on observations obtained at the W. M. Keck Observatory, which is operated jointly by the University of California and the California Institute of Technology. Keck time was granted for this project by the Caltech, the University of Hawai‘i, the University of California, and NASA.}}

\correspondingauthor{Dimitri Mawet}
\email{dmawet@astro.caltech.edu}

\author{Dimitri Mawet}
\affiliation{Department of Astronomy, California Institute of Technology, Pasadena, CA 91125, USA}
\affiliation{Jet Propulsion Laboratory, California Institute of Technology, Pasadena, CA 91109, USA}

\author{Lea Hirsch}
\affiliation{Kavli Institute for Particle Astrophysics and Cosmology, Stanford University, Stanford, CA 94305, USA}
\affiliation{University of California, Berkeley, 510 Campbell Hall, Astronomy Department, Berkeley, CA 94720, USA}

\author{Eve J. Lee}
\affiliation{TAPIR, Walter Burke Institute for Theoretical Physics, Mailcode 350-17, Caltech, Pasadena, CA 91125, USA}

\author{Jean-Baptiste Ruffio}
\affiliation{Kavli Institute for Particle Astrophysics and Cosmology, Stanford University, Stanford, CA, 94305, USA}

\author{Michael Bottom}
\affiliation{Jet Propulsion Laboratory, California Institute of Technology, Pasadena, CA 91109, USA}

\author{Benjamin J. Fulton}
\affiliation{NASA Exoplanet Science Institute, Caltech/IPAC-NExScI, 1200 East California Boulevard, Pasadena, CA 91125, USA}

\author{Olivier Absil}
\affiliation{Space sciences, Technologies \& Astrophysics Research (STAR) Institute, Universit\'e de Li\`ege, All\'ee du Six Ao\^ut 19c, B-4000 Sart Tilman, Belgium}
\affiliation{F.R.S.-FNRS Research Associate}

\author{Charles Beichman}
\affiliation{Jet Propulsion Laboratory, California Institute of Technology, Pasadena, CA 91109, USA}
\affiliation{Division of Physics, Mathematics, and Astronomy, California Institute of Technology, Pasadena, CA 91125, USA}
\affiliation{NASA Exoplanet Science Institute, 770 S. Wilson Avenue, Pasadena, CA 911225, USA}

\author{Brendan Bowler}
\affiliation{McDonald Observatory and the University of Texas at Austin, Department of Astronomy, 2515 Speedway, Stop C1400, Austin, TX 78712, USA}

\author{Marta Bryan}
\affiliation{Department of Astronomy, California Institute of Technology, Pasadena, CA 91125, USA}

\author{Elodie Choquet}
\affiliation{Department of Astronomy, California Institute of Technology, Pasadena, CA 91125, USA}
\affiliation{Hubble Postdoctoral Fellow}

\author{David Ciardi}
\affiliation{NASA Exoplanet Science Institute, Caltech/IPAC-NExScI, 1200 East California Boulevard, Pasadena, CA 91125, USA}

\author{Valentin Christiaens}
\affiliation{Space sciences, Technologies \& Astrophysics Research (STAR) Institute, Universit\'e de Li\`ege, All\'ee du Six Ao\^ut 19c, B-4000 Sart Tilman, Belgium}
\affiliation{Departamento de Astronom\'ia, Universidad de Chile, Casilla 36-D, Santiago, Chile}
\affiliation{Millenium Nucleus ``Protoplanetary Disks in ALMA Early Science'', Chile}

\author{Denis Defr\`ere}
\affiliation{Space sciences, Technologies \& Astrophysics Research (STAR) Institute, Universit\'e de Li\`ege, All\'ee du Six Ao\^ut 19c, B-4000 Sart Tilman, Belgium}

\author{Carlos Alberto Gomez Gonzalez}
\affiliation{Universit\'e Grenoble Alpes, IPAG, F-38000 Grenoble, France}

\author{Andrew W. Howard}
\affiliation{Department of Astronomy, California Institute of Technology, Pasadena, CA 91125, USA}

\author{Elsa Huby}
\affiliation{LESIA, Observatoire de Paris, Meudon, France}

\author{Howard Isaacson}
\affiliation{University of California, Berkeley, 510 Campbell Hall, Astronomy Department, Berkeley, CA 94720, USA}

\author{Rebecca Jensen-Clem}
\affiliation{University of California, Berkeley, 510 Campbell Hall, Astronomy Department, Berkeley, CA 94720, USA}
\affiliation{Miller Fellow}

\author{Molly Kosiarek}
\affiliation{University of California, Santa Cruz; Department of Astronomy and Astrophysics, Santa Cruz, CA 95064, USA}
\affiliation{NSF Graduate Research Fellow}

\author{Geoff Marcy}
\affiliation{University of California, Berkeley, 510 Campbell Hall, Astronomy Department, Berkeley, CA 94720, USA}

\author{Tiffany Meshkat}
\affiliation{IPAC, California Institute of Technology, M/C 100-22, 770 S. Wilson Ave, Pasadena, CA 91125, USA}

\author{Erik Petigura}
\affiliation{Department of Astronomy, California Institute of Technology, Pasadena, CA 91125, USA}

\author{Maddalena Reggiani}
\affiliation{Space sciences, Technologies \& Astrophysics Research (STAR) Institute, Universit\'e de L\`ege, All\'ee du Six Ao\^ut 19c, B-4000 Sart Tilman, Belgium}

\author{Garreth Ruane}
\affiliation{Department of Astronomy, California Institute of Technology, Pasadena, CA 91125, USA}
\affiliation{NSF Astronomy and Astrophysics Postdoctoral Fellow}

\author{Eugene Serabyn}
\affiliation{Jet Propulsion Laboratory, California Institute of Technology, Pasadena, CA 91109, USA}

\author{Evan Sinukoff}
\affiliation{Institute for Astronomy, University of Hawai‘i at Manoa, Honolulu, HI 96822, USA}

\author{Ji Wang}
\affiliation{Department of Astronomy, California Institute of Technology, Pasadena, CA 91125, USA}

\author{Lauren Weiss}
\affiliation{Institut de Recherche sur les Exoplanètes, Dèpartement de Physique, Universitè de Montrèal,
C.P. 6128, Succ. Centre-ville, Montréal, QC H3C 3J7, Canada}

\author{Marie Ygouf}
\affiliation{IPAC, California Institute of Technology, M/C 100-22, 770 S. Wilson Ave, Pasadena, CA 91125, USA}

\begin{abstract}

We present the most sensitive direct imaging and radial velocity (RV) exploration of $\epsilon$ Eridani to date. $\epsilon$ Eridani is an adolescent planetary system, reminiscent of the early Solar system. It is surrounded by a prominent and complex debris disk which is likely stirred by one or several gas giant exoplanets. The discovery of the RV signature of a giant exoplanet was announced 15 years ago, but has met with scrutiny due to possible confusion with stellar noise. We confirm the planet with a new compilation and analysis of precise RV data spanning 30 years, and combine it with upper limits from our direct imaging search, the most sensitive ever performed. The deep images were taken in the Ms band (4.7$\mu$m) with the vortex coronagraph recently installed in W.M. Keck Observatory's infrared camera NIRC2, which opens a sensitive window for planet searches around nearby adolescent systems. The RV data and direct imaging upper limit maps were combined in an innovative joint Bayesian analysis, providing new constraints on the mass and orbital parameters of the elusive planet. $\epsilon$ Eridani b has a mass of $0.78^{+0.38}_{-0.12}$ $M_{Jup}$ and is orbiting $\epsilon$ Eridani at about $3.48\pm 0.02$ AU with a period of $7.37 \pm 0.07$ years. The eccentricity of $\epsilon$ Eridani b's orbit is $0.07^{+0.06}_{-0.05}$, an order of magnitude smaller than early estimates and consistent with a circular orbit. We discuss our findings from the standpoint of planet-disk interactions and prospects for future detection and characterization with the James Webb Space Telescope.

\end{abstract}


\section{Introduction} \label{sec:intro}

\thisstar\ is an adolescent \citep[200-800 Myr,][]{Fuhrmann2004,Mamajek2008} K2V dwarf star (Table \ref{table1}). At a distance of 3.2 pc, \thisstar\ is the 10th closest star to the Sun, which makes it a particularly attractive target for deep planet searches. Its age, spectral type, distance and thus apparent brightness ($V=3.73$ mag.) make it a benchmark  system, as well as an excellent analog for the early phases of the Solar system's evolution. \thisstar\ hosts a prominent, complex debris disk, and a putative Jupiter-like planet.

\subsection{\thisstar's debris disk}

The disk was first detected by the Infrared Astronomical Satellite \citep[IRAS, ][]{Aumann1985} and later on by the Infrared Space Observatory \citep[ISO, ][]{Walker2000}. It was first imaged by the Submillimetre Common-User Bolometer Array at the James Clerk Maxwell Telescope by \citet{Greaves1998}. \thisstar\ is one of the ``fabulous four'' Vega-like debris disks and shows more than 1 Jy of far infrared (FIR) excess over the stellar photosphere at 60-200 $\mu$m and a lower significance excess at 25 $\mu$m \citep{Aumann1985}. Using the Spitzer Space Telescope and the Caltech Submillimeter Observatory (CSO) to trace \thisstar's spectral energy distribution (SED) from 3.5 to 350 $\mu$m, the model presented in \citet{Backman2009} paints a complex picture. \thisstar's debris disk is composed of a main ring at 35-90 AU, and a set of two narrow inner dust belts inside the cavity delineated by the outer ring: one belt with a color temperature $T\simeq 55$ K at approximately 20 AU, and another belt with a color temperature $T\simeq 120$ K at approximately 3 AU \citep{Backman2009}. The authors argue that to maintain the three-belt system around $\epsilon$ Eridani, three shepherding planets are necessary.

Using Herschel at 70, 160, 250, 350, and 500 $\mu$m, \citet{Greaves2014} refined the position and the width of the outer belt to be 54-68 AU, but only resolved one of the inner belts at 12-16 AU. More recently, \citet{Chavez2016} used the Large Millimetre Telescope Alfonso Serrano (LMT) at 1.1 mm and resolved the outer belt at a separation of 64 AU. Emission is detected at the location of the star in excess of the photosphere. The angular resolution of the 1.1 mm map is however not sufficient to resolve the inner two warm belts of \citet{Backman2009}. \citet{MacGregor2015} used the Submillimeter Array (SMA) at 1.3 mm and the Australia Telescope Compact Array at 7 mm to resolve the outer ring, and measure its width (measurement now superseded by ALMA, see below). The data at both mm wavelengths show excess emission, which the authors attribute to ionized plasma from a stellar corona or chromosphere.

\citet{Su2017} recently presented 35 $\mu$m images of \thisstar\ obtained with the Stratospheric Observatory for Infrared Astronomy (SOFIA). The inner disk system is marginally resolved within 25 AU. Combining the $15-38$ $\mu$m excess spectrum with Spitzer data, \citet{Su2017} find that the presence of in situ dust-producing planetesimal belt(s) is the most likely source of the infrared excess emission in the inner 25 AU region. However, the SOFIA data are not constraining enough to distinguish one broad inner disk from two narrow belts. 

\citet{Booth2017} used the Atacama Large Millimeter/submillimeter Array (ALMA) to image the Northern arc of the outer ring at high angular resolution (beam size $<2\arcsec$). The 1.34-mm continuum image is low signal-to-noise but well resolved, with the outer ring extending from 62.6 to 75.9 AU. The fractional outer disk width is comparable to that of the Solar system's Kuiper Belt and makes it one of the narrowest debris disks known, with a width of just $\simeq 12$ AU. The outer ring inclination is measured to be $i = 34^{\circ}\pm2^{\circ}$, consistent with all previous estimates using lower resolution sub-mm facilities \citep{Greaves1998,Greaves2005,Backman2009}. No significant emission is detected between $\sim$20 and $\sim$60 AU \citep[see][their Figure~5]{Booth2017}, suggesting a large clearing between the inner belt(s) within 20 AU and the outer belt outward of 60 AU. \citet{Booth2017} find tentative evidence for clumps in the ring, and claim that the inner and outer edges are defined by resonances with a planet at a semi-major axis of 48 AU. The authors also confirm the previous detection of unresolved mm emission at the location of the star that is above the level of the photosphere and attribute this excess to stellar chromospheric emission, as suggested by \citet{MacGregor2015}. However, the chromospheric emission cannot reproduce the infrared excess seen by Spitzer and SOFIA. 

Finally, recent 11-$\mu$m observations with the Large Binocular Telescope Interferometer (LBTI) suggest that warm dust is present within $\sim$500\,mas (or 1.6\,AU) from \thisstar\ \citep{Ertel2018}. The trend of the detected signal with respect to the stellocentric distance also indicates that the bulk of the emission comes from the outer part of the LBTI field-of-view and, hence, likely associated to the dust belt(s) responsible from the 15-38~$\mu$m emission detected by Spitzer and SOFIA. 

\subsection{\thisstar's putative planet}

\citet{Hatzes2000} demonstrated that the most likely explanation for the observed decade-long radial velocity (RV) variations was the presence of a $\simeq 1.5$ M$_J$ giant planet with a period $P=6.9$ yr ($\simeq 3$ AU orbit) and a high eccentricity ($e=0.6$). While most of the exoplanet community seems to have acknowledged the existence of $\epsilon$ Eridani b, there is still a possibility that the measured RV variations are due to stellar activity cycles \citep{Anglada2012,Zechmeister2013}. \citet{Backman2009} rightfully noted that a giant planet with this orbit would quickly clear the inner region not only of dust particles but also the parent planetesimal belt needed to resupply them, inconsistent with their observations.

\subsection{This paper}

In this paper, we present the deepest direct imaging reconnaissance of \thisstar\ to date, a compilation of precision radial velocity measurements spanning 30 years, and an innovative joint Bayesian analysis combining both planet detection methods. Our results place the tightest constraints yet on the planetary mass and orbital parameters for this intriguing planetary system. The paper is organized as follows: \sectionautorefname~\ref{sec:doppler} describes the RV observations and data analysis, \sectionautorefname~\ref{sec:hci} describes the high contrast imaging observations and post-processing, \sectionautorefname~\ref{sec:results} presents our nondetection and robust detection limits from direct imaging, additional tests on the RV data, as well as our joint analysis of both data sets. In \sectionautorefname~\ref{sec:discussion}, we discuss our findings, the consequences of the new planet parameters on planet-disk interactions, and prospects for detection with the James Webb Space Telescope, before concluding in \sectionautorefname~\ref{sec:conclusions}.

\begin{deluxetable}{ccc}
\tabletypesize{\scriptsize}

\tablecaption{Properties of \thisstar \label{table1}}
\tablewidth{0pt}
\tablehead{
\colhead{Property} & \colhead{Value} & \colhead{Ref.}}
\startdata
RA (hms) & 03 32 55.8 (J2000) &\citet{vanLeeuwen2007}\\
DEC (dms) & -09 27 29.7 (J2000) &\citet{vanLeeuwen2007}\\
Spect. type & K2V &\citet{Keenan1989}\\
Mass ($M_\odot$) & $0.781 \pm 0.005$ &\citet{Boyajian2012}\\
Distance (pc) & $3.216 \pm 0.0015$ &\citet{vanLeeuwen2007}\\
$V$ mag. & 3.73 &\citet{Ducati2002} \\ 
$K$ mag. & 1.67 &\citet{Ducati2002}\\
$L$ mag. & 1.60 &\citet{Cox2000}\\
$M$ mag. & 1.69 &\citet{Cox2000}\\
Age (Myr) &200-800 &\citet{Mamajek2008}\\
\enddata
\end{deluxetable}

\section{Doppler spectroscopy} \label{sec:doppler}

In this section, we present our new compilation and analysis of Doppler velocimetry data of \thisstar\ spanning 30 years. 

\subsection{RV observations} \label{subsec:doppler_obs}

\thisstar\ has been included in planet search programs at both Keck Observatory using the HIRES Spectrometer \citep{Howard2010} and at Lick Observatory using the Automated Planet Finder (APF) and Levy Spectrometer. \thisstar\ was observed on 206 separate nights with HIRES and the APF, over the past 7 years. 

Keck/HIRES \citep{Vogt1994} radial velocity observations were obtained starting in 2010, using the standard iodine cell configuration of the California Planet Survey \citep[CPS,][]{Howard2010}. During the subsequent 7 years, 91 observations were taken through the B5 or C2 deckers ($0\farcs87\times3\farcs5$ and $0\farcs87\times14\arcsec$ respectively), yielding a spectral resolution of $R \approx 55,000$ for each observation. Each measurement was taken through a cell of gaseous molecular iodine heated to $50^{\circ}$C, which imprints a dense forest of iodine absorption lines onto the stellar spectrum in the spectral region of $5000 - 6200 \AA$. This iodine spectrum was used for wavelength calibration and as a PSF reference. Each RV exposure was timed to yield a per-pixel SNR of $\gtrsim 200$ at 550 nm, with typical exposure times of only a few seconds due to the brightness of the target. An iodine-free template spectrum was obtained using the B3 decker ($0\farcs57\times14\arcsec$, $R\approx72,000$) on 30 August, 2010. 

RV observations using the Automated Planet Finder (APF) and Levy Spectrograph \citep{Radovan2014,Vogt2014} were taken starting in late 2013. The APF is a 2.4-m telescope dedicated to performing radial velocity detection and follow-up of planets and planet candidates, also using the iodine cell method of wavelength calibration. APF data on \thisstar\ were primarily taken through the W decker ($1\arcsec \times 3\arcsec$), with a spectral resolution of $R \approx 110,000$. Exposures were typically between 10 and 50 seconds long, yielding SNR per-pixel of 140. The typical observing strategy at the APF was to take three consecutive exposures and then bin them to average over short-term fluctuations from stellar oscillations. In some nights, more than one triple-exposure was taken. These were binned on a nightly timescale. An iodine-free template consisting of five consecutive exposures was obtained on 17 February, 2014 using the N decker ($0\farcs5\times8\arcsec$) with resolution $R\approx150,000$. Both the APF and HIRES RVs were calibrated to the solar system barycenter and corrected for the changing perspective caused by the high proper motion of \thisstar.

In addition to the new HIRES and APF radial velocity data, we incorporate previously published data from several telescopes into this study. High precision radial velocity observations of \thisstar\ were taken with the Hamilton spectrograph at Lick Observatory starting in 1987, as part of the Lick Planet Search program. They were published in the catalog of \citet{Fischer2014}, along with details of the instrumental setup and reduction procedure. 

The Coud\'e Echelle Spectrograph at La Silla Observatory was used, first with the Long Camera on the 1.4-m telescope from 1992-1998, then with the Very Long Camera on the 3.6-m telescope from 1998-2006, to collect additional radial velocity data  on \thisstar. RV data were also collected using the HARPS spectrograph, also on the 3.6-m telescope at La Silla Observatory, during 2004-2008. Together, these data sets are published in \cite{Zechmeister2013}.

\subsection{RV data analysis}\label{subsec:doppler_analysis}

All new spectroscopic observations from Keck/HIRES and APF/Levy were reduced using the standard CPS pipeline \citep{Howard2010}. The iodine-free template spectrum was deconvolved with the instrumental PSF and used to forward model each observation's relative radial velocity. The iodine lines imprinted on the stellar spectrum by the iodine cell were used as a stable wavelength calibration, and the instrumental PSF was modeled as the sum of several Gaussians \citep{Butler1996}. Per the standard CPS RV pipeline, each spectrum was divided into approximately 700 spectral ``chunks'' for which the radial velocity was individually calculated. The final radial velocity and internal precision were calculated as the weighted average of each of these chunks. Chunks with RVs that are more consistent across observations are given higher weight. Those chunks with a larger scatter are given a lower weight. The RV observations from Keck/HIRES and APF/Levy are listed in Tables \ref{tab:rvs_k} and \ref{tab:rvs_a} respectively. These data are not offset-subtracted to account for different zero-points, and the uncertainties reported in the Tables reflect the weighted standard deviations of the chunk-by-chunk radial velocities, and do not include systematic uncertainty such as jitter. 

In combination among HIRES, the APF, and the other instruments incorporated, a total of 458 high-precision radial velocity observations have been taken, over an unprecedented time baseline of 30 years. We note that the literature radial velocity data included in this study were analyzed using a separate radial velocity pipeline, which in some cases did not include a correction for the secular acceleration of the star. Secular acceleration is caused by the space motion of a star, and depends on its proper motion and distance. For \thisstar, we calculate a secular acceleration signal of $0.07\ \mathrm{m s^{-1} yr^{-1}}$ \citep{Zechmeister2009}. This is accounted for in the HIRES and APF data extraction, and most likely in the Lick pipeline, but is not included in the reduction for the CES and HARPS data. However, the amplitude of this effect is much smaller than the amplitude of RV variation observed in the radial velocities due to stellar activity and the planetary orbit. Since each of the CES and HARPS data sets cover less than 10 years, we expect to see less than 1 \ms\ variation across each full data set. We tested running the analysis with and without applying these corrections. All of the resulting orbital parameters were consistent to within $\sim 0.1 \sigma$. We therefore neglect this correction in our radial velocity analyses. From the combined radial velocity data set, a clear periodicity of approximately 7 years is evident, both by eye and in a periodogram of the RV data. We assess this periodicity in \sectionautorefname~\ref{subsec:robustness}. 

\thisstar's youth results in significant stellar magnetic activity. Convection-induced motions on the stellar surface cause slight variations in the spectral line profiles, leading to variations in the inferred radial velocity that do not reflect motion caused by a planetary companion. As a result, stellar magnetic activity may mimic the radial velocity signal of an orbiting planet, resulting in false positives. We therefore extract \shk\ values from each of the HIRES and APF spectra taken for \thisstar. \shk\ is a measure of the excess emission at the cores of the Ca II H \& K lines due to chromospheric activity, and correlates with stellar magnetic activity such as spots and faculae which might have effects on the radial velocities extracted from the spectra \citep{Isaacson2010}.

\section{High contrast imaging} \label{sec:hci}

Here, we present our new deep, direct, high contrast imaging observations and data analysis of \thisstar\ using the Keck NIRC2 vortex coronagraph.

\subsection{High contrast imaging observations}\label{subsec:hci_obs}

We observed \thisstar\ over three consecutive nights in January 2017 (see Table~\ref{table2}). We used the vector vortex coronagraph installed in NIRC2 \citep{Serabyn2017}, the near-infrared camera and spectrograph behind the adaptive optics system of the 10-m Keck II telescope at W.M. Keck Observatory. The vortex coronagraph is a phase-mask coronagraph enabling high contrast imaging at very small angles close to the diffraction limit of the 10-meter Keck telescope at $4.67 \mu$m ($\simeq 0\farcs1$). The starlight suppression capability of the vortex coronagraph is induced by a $4\pi$ radian phase ramp wrapping around the optical axis. When the coherent adaptively-corrected point spread function (PSF) is centered on the vortex phase singularity, the on-axis starlight is redirected outside the geometric image of the telescope pupil formed downstream from the coronagraph, where it is blocked by means of an undersized diaphragm (the Lyot stop). The vector vortex coronagraph installed in NIRC2 was made from a circularly concentric subwavelength grating etched onto a synthetic diamond substrate \citep[Annular Groove Phase Mask coronagraph or AGPM,][]{Mawet2005, Vargas2016}. 

\begin{deluxetable*}{cccc}
\tabletypesize{\scriptsize}

\tablecaption{Observing log for NIRC2 imaging data\label{table2}}
\tablewidth{0pt}
\tablehead{
\colhead{Properties} & \colhead{value} & \colhead{value} & \colhead{value}}
\startdata
UT date (yyyy-mm-dd) & 2017-01-09 & 2017-01-10 & 2017-01-11 \\
UT start time (hh:mm:ss) & 05:11:55 & 05:12:47 & 05:48:08 \\
UT end time (hh:mm:ss) & 09:14:11 & 09:31:09 & 08:36:14 \\
Discr. Int. Time (s) & 0.5 & - & - \\
Coadds & 60 & - & - \\
Number of frames & 210 & 260 & 154\\
Total integration time (s) & 6300 & 7800 & 4620\\
Plate scale (mas/pix) & 9.942 (``narrow'') & - & -\\
Total FoV & $r\simeq 5\arcsec$ (vortex mount) & - & -\\
Filter &Ms $[4.549,4.790]$ $\mu$m &- &-\\
Coronagraph &Vortex (AGPM) &- &-\\ 
Lyot stop &Inscribed circle & - & -\\
0.5-$\mu$m DIMM seeing ($\arcsec$) &0.52 &0.64 &0.97\\
Par. angle start-end ($^\circ$) &-36 -- +52 &-35 -- +55 &-20 -- +46\\
\enddata
\end{deluxetable*}

Median 0.5-$\mu$m DIMM seeing conditions ranged from $0\farcs 52$ to $0\farcs 97$ (see Table \ref{table2}). The adaptive optics system provided excellent correction in the Ms-band 
($[4.549,4.790]$ $\mu$m) with Strehl ratio of about $90\%$ (NIRC2 quicklook estimate), similar to the image quality provided at shorter wavelengths by extreme adaptive optics systems such as the Gemini Planet Imager \citep[GPI,][]{Macintosh2014}, SPHERE \citep{Beuzit2008}, and SCExAO \citep{Jovanovic2015}. The alignment of the star onto the coronagraph center, a key to high contrast at small angles, was performed using the quadrant analysis of coronagraphic images for tip-tilt sensing \citep[QACITS,][]{Huby2015,Huby2017}. The QACITS pointing control uses NIRC2 focal-plane coronagraphic science images in a closed feedback loop with the Keck adaptive optics tip-tilt mirror \citep{Serabyn2017,Mawet2017,Huby2017}. The typical low-frequency centering accuracy provided by QACITS is $\simeq 0.025\lambda/D$ rms, or $\simeq 2$ mas rms. 

All of our observations were performed in vertical angle mode, which forces the AO derotator to track the telescope pupil (following the elevation angle) instead of the sky, effectively allowing the field to rotate with the parallactic angle, enabling angular differential imaging \citep[ADI,][]{Marois2006}. 

\subsection{Image post-processing} \label{subsec:hci_analysis}

After correcting for bad pixels, flat-fielding, subtracting sky background frames using principal component analysis (PCA), and co-registering the images, we applied PCA \citep{Soummer2012,Gomez2017} to estimate and subtract the post-coronagraphic residual stellar contribution from the images. We used the open-source Vortex Image Processing -- \VIP \footnote{https://github.com/vortex-exoplanet/VIP} -- software package \citep{Gomez2017}, and applied PCA on the combined data from all 3 nights, totaling more than 5 hours of open shutter integration time (see Table~\ref{table2} for details). We used a numerical mask $2\lambda/D$ in radius to occult the bright stellar residuals close to the vortex coronagraph inner working angle.

\begin{figure}
    \includegraphics[width=8.5cm]{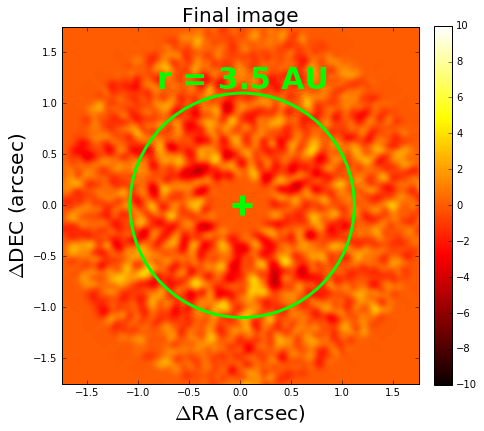}
    \caption{Final reduced image of \thisstar, using PCA, and 120 principal components in the PSF reconstruction. The scale is linear in analog to digital units (ADU).}
    \label{fig:finalimage}
\end{figure}

The final image (Figure~\ref{fig:finalimage}) was obtained by pooling all three nights together in a single dataset totaling 624 frames. The PSF was reconstructed by using 120 principal components and projections on the $351\times 351$ pixel frames excluding the central numerical mask ($2\lambda/D$ in radius). This number of principal components was optimized to yield the best final contrast limits in the 1-5 AU region of interest, optimally trading-off speckle noise and self-subtraction effects. The final image (Figure~\ref{fig:finalimage}) does not show any particular feature and is consistent with whitened speckle noise.

\section{Analysis} \label{sec:results}

In this section, we present our nondetection and robust detection limits from direct imaging, additional tests on the RV data, as well as our joint analysis of both data sets.

\subsection{Robust detection limits from direct imaging} \label{subsec:results1}

Following \citet{Mawet2014}, we assume that ADI and PCA post-processing whiten the residual noise in the final reduced image through two complementary mechanisms. First, PCA removes the correlated component of the noise by subtracting off the stellar contribution, revealing underlying independent noise processes such as background, photon Poisson noise, readout noise, and dark current. Second, the ADI frame combination provides additional whitening due to the field rotation during the observing sequence and subsequent derotation, and by virtue of the central limit theorem, regardless of the underlying distribution of the noise \citep{Marois2006,Marois2008}. From now on, we assume Gaussian statistics to describe the noise of our images. Our next task is to look for point sources, and if none are found, place meaningful upper limits. Whether or not point sources are found, we will use our data to constrain the planet mass posterior distribution as a function of projected separation.

For this task, we choose to convert flux levels into mass estimates using the COND evolutionary model \citep{Baraffe2003} for the three ages considered in this work: 200, 400, and 800 Myr. The young age end of our bracket (200 Myr) is derived from a pure kinematic analysis \citep{Fuhrmann2004}. The 400 and 800 Myr estimates are from \citet{Mamajek2008}, who used chromospheric activities and spin as age indicators.

As noted by \citet{Bowler2016}, the COND model is part of the hot-start model family, which begins with arbitrarily large radii and oversimplified, idealized initial conditions. It ignores the effects of accretion and mass assembly. The COND model represents the most luminous and thus optimistic outcome. At the adolescent age range of \thisstar, initial conditions of the formation of a Jupiter-mass gas giant have mostly been forgotten \citep{Marley2007,fortney2008} and have a minor impact on mass estimates. Moreover, a very practical reason why COND was used is because it is the only model readily providing open-source tables extending into the low-mass regime ($<1 M_{Jup}$) reached by our data (see \S \ref{sub:models}).

\subsubsection{Direct imaging nondetection}\label{subsub:SD}

Signal detection is a balancing act where one trades off the risk of false alarm with sensitivity. The signal detection threshold $\tau$ is related to the risk of false alarm or false positive fraction (FPF) as follows 

\begin{equation}\label{eq1}
FPF=\frac{FP}{TN+FP}=\int_{\tau}^{+\infty} p(x | H_0) dx
\end{equation}
where $x$ is the intensity of the residual speckles in our images, and $p(x | H_0)$, the probability density function of $x$ under the null hypothesis $H_0$. FP is the number of false positives and TN, the number of true negatives. Assuming Gaussian noise statistics, the traditional $\tau=5\sigma$ threshold yields $2.98\times 10^{-7}$ false alarm probability, or FPF. 

\begin{figure}
    \includegraphics[width=8.5cm]{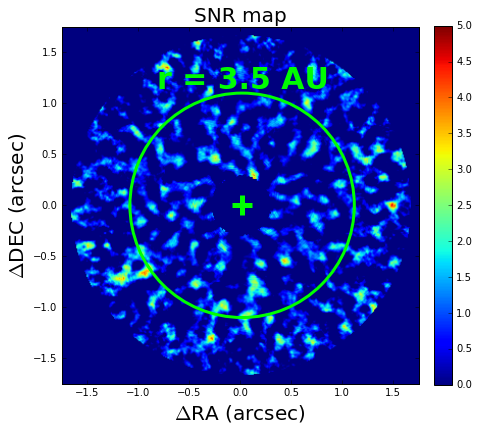}
    \caption{SNR map for our most sensitive reduction, using 120 principal components. We used the SNR map function implemented in open source package VIP \citep{Gomez2017}. The method uses the annulus-wise approach presented in \citet{Mawet2014}. No source is detected above $5\sigma$. The green circle delineates the planet's project separation at $\simeq 3.5$ AU.}
    \label{fig:snrmap}
\end{figure}

Applying the $\tau=5\sigma$ threshold to the signal-to-noise ratio (SNR) map generated from our most sensitive reduction, which occurs for a number of principal components equal to 120, yields no detection, consistent with a null result. In other words, \thisstar\ b is not detected in our deep imaging data to the $5\sigma$ threshold. To compute the SNR map, we used the annulus-wise approach outlined in \citet{Mawet2014}, and implemented in the open-source python-based Vortex Imaging Pipeline \citep{Gomez2017}. The noise in an annulus at radius $r$ (units of $\lambda/D$) is computed as the standard deviation of the $n=2\pi r$ resolution elements at that radius. The algorithm throughput is computed using fake companion injection-recovery tests at every location in the image. This step is necessary to account for ADI self-subtraction effects. The result is shown in Figure~\ref{fig:snrmap}. 

To quantify our sensitivity, also known as ``completeness'', we use the true positive fraction (TPF), defined as

\begin{equation}\label{eq2}
TPF=\frac{TP}{TP+FN}=\int_{\tau}^{+\infty} p(x | H_1) dx
\end{equation}
with $p(x | H_1)$, the probability density function of $x$ under the hypothesis $H_1$ - signal present, and where TP is the number of true positives and FN, the number of false negatives. For instance, a $95\%$  sensitivity (or completeness) for a given signal $I$, and detection threshold $\tau$ means that $95\%$ of the objects at the intensity level $I$ will statistically be recovered from the data. 

\begin{figure}
    \includegraphics[width=8.5cm]{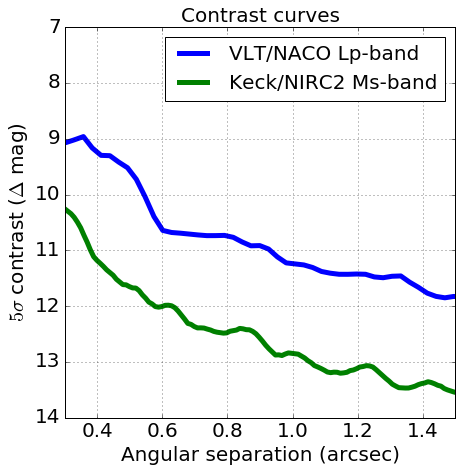}
    \caption{Traditional $\tau=5\sigma$ contrast curves comparing our Keck/NIRC2 vortex coronagraph Ms-band data and the VLT/NACO Lp-band data (PI: Quanz, Program ID: 090.C-0777(A)) presented in \citet{Mizuki2016}, and reprocessed here with the VIP package.}
    \label{fig:contrast}
\end{figure}

\begin{figure*}
    \centering
    \includegraphics[width=\textwidth]{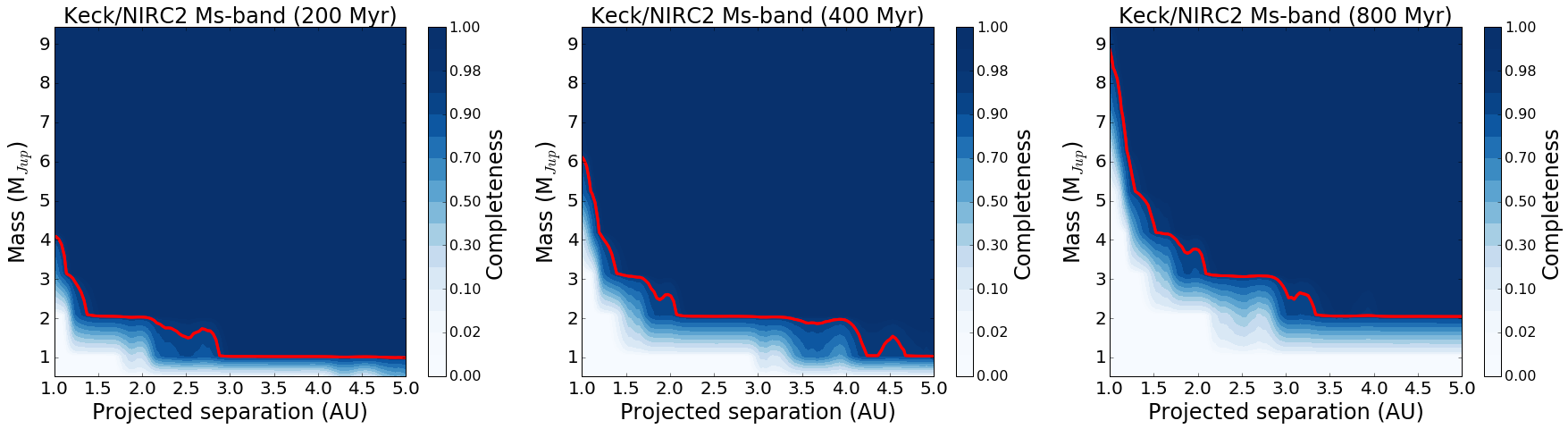}
    \caption{Keck/NIRC2 Ms-band vortex performance/completeness maps for a $\tau=5\sigma$ detection threshold for all 3 different ages considered here. The red curve highlights the 95\% completeness contour.}
    \label{fig:sensitivity_nirc2}
\end{figure*}

The sensitivity contours, or ``performance maps'' \citep{Jensen-Clem2018} for a uniform threshold corresponding to $2.98\times 10^{-7}$ false alarm probability (FPF) are shown in Figure~\ref{fig:sensitivity_nirc2}. The choice of threshold is assuming Gaussian noise statistics and accounts for small sample statistics as in \citet{Mawet2014}. At the location of the elusive RV exoplanet, the threshold corrected for small sample statistics converges to $\tau \approx 5\sigma$. The corresponding traditional $\tau=5\sigma$ contrast curve at 50\% completeness is shown in Figure~\ref{fig:contrast}.

\begin{figure*}
    \centering
    \includegraphics[width=\textwidth]{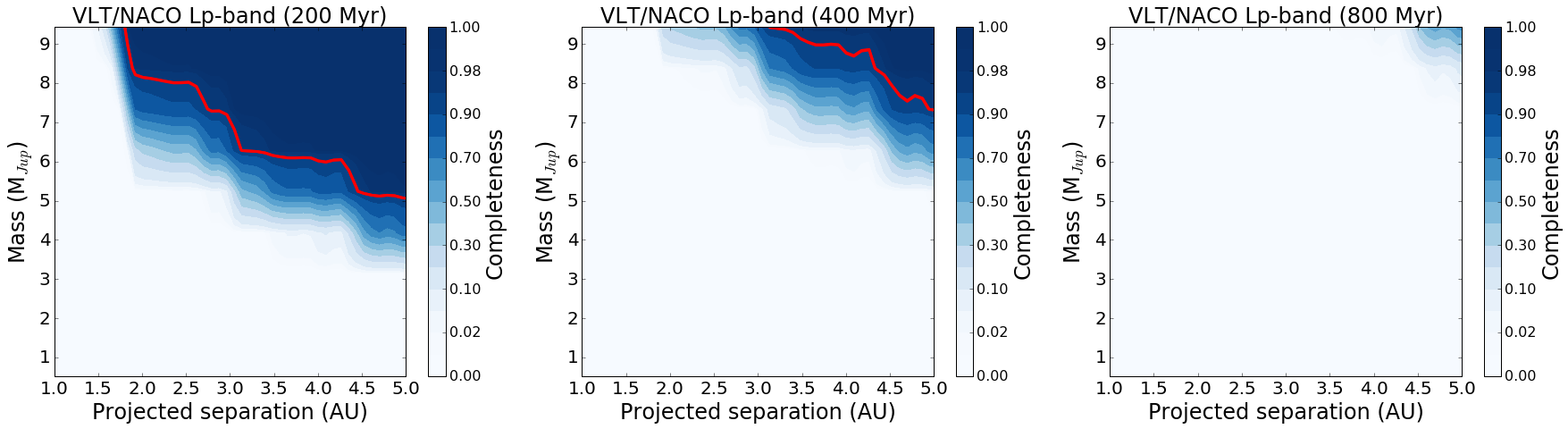}
    \caption{Performance/completeness maps for a $\tau=5\sigma$ detection threshold for all 3 different ages considered here using VLT/NACO Lp-band data (PI: Quanz, Program ID: 090.C-0777(A)) presented in \citet{Mizuki2016}. The red curve highlights the 95\% completeness contour.}
    \label{fig:sensitivity_naco}
\end{figure*}

\subsubsection{Comparison to previous direct imaging results}

\citet{Mizuki2016} presented an extensive direct imaging compilation and data analysis for \thisstar. The authors analyzed data from Subaru/HiCIAO, Gemini/NICI, and VLT/NACO. Here we focus on the deepest data set reported in \citet{Mizuki2016}, which is the Lp-band NACO data from PI: Quanz (Program ID: 090.C-0777(A)). This non-coronagraphic ADI sequence totals 146.3 minutes of integration time and about 67 degrees of parallactic angle rotation. \citet{Mizuki2016}'s reported $5\sigma$ and 50\% completeness mass sensitivity using the hot start COND evolutionary model is $> 10M_J$ at 1 AU for all 3 ages considered here, i.e., 200, 400, and 800 Myr. At 2 AU, it is $\simeq 2.5M_J$, $\simeq 4M_J$, and $\simeq 6.5M_J$, respectively. At 3 AU, it is $\simeq 2M_J$, $\simeq 3M_J$, and $\simeq 5M_J$, respectively. 

For consistency, we reprocessed the VLT/NACO data with the VIP package and computed completeness maps using the same standards as for our Keck/NIRC2 data. The results are shown in Figure~\ref{fig:sensitivity_naco}. Our computed $5\sigma$ and 50\% completeness mass sensitivity using the hot start COND evolutionary model for the VLT/NACO Lp-band data is $> 10M_J$ at 1 AU for all 3 ages considered here, i.e., 200, 400, and 800 Myr. At 2 AU, it is $\simeq 6.5M_J$ at 200 Myr and $> 10 M_J$ at both 400 and 800 Myr. At 3 AU, it is $\simeq 5.5 M_J$, $\simeq 8 M_J$, and $> 10 M_J$, respectively. 

Our computed $5\sigma$ and 50\% completeness results for the VLT/NACO Lp-band data are systematically worse than those presented in \citet{Mizuki2016}. We note a factor 2 discrepancy in mass between the published results and our values. We suggest that it may be the result of inaccurate flux loss calibrations in \citet{Mizuki2016}, which is a common occurrence with ADI data sets. 

We find that our Ms-band Keck/NIRC2 coronagraphic data is about a factor 5-10 more sensitive in mass across the range of Solar-system scales probed in this work than the previously best available data set. Our $5\sigma$ and 50\% completeness mass sensitivity using the hot start COND evolutionary model is $\simeq 3 M_J$, $\simeq 4.5 M_J$, and $\simeq 6.5 M_J$ at 1 AU for all 3 ages considered here, i.e., 200, 400, and 800 Myr, respectively. At 2 AU, it is $\simeq 1.5M_J$, $\simeq 1.7 M_J$, and $\simeq 2.5M_J$, respectively. At 3 AU, it is $\simeq 0.8 M_J$, $\simeq 1.7M_J$, and $\simeq 5M_J$, respectively. 

These result demonstrates the power of ground-based Ms-band small-angle coronagraphic imaging for nearby adolescent systems. When giant exoplanets cool down to below 1000 K, the peak of their black body emission shifts to 3-5 $\mu$m mid-infrared wavelengths. Moreover, due to the $t^{-5/4}$ dependence of bolometric luminosity on age \citep{Stevenson1991}, mid-infrared luminosity stays relatively constant for hundreds of millions of years.

\subsection{Tests on the RV Data}
\label{subsec:robustness}

In light of our nondetection of a planet in the NIRC2 high-contrast imaging, we consider the possibility that the planet is not real, or that the periodicity is caused by stellar activity. We utilize the RV analysis package \radvel\footnote{Documentation available at \url{http://radvel.readthedocs.io/en/latest/}} \citep{Fulton2018} to perform a series of tests to determine the significance of the periodicity and attempt to rule out stellar activity as its source. We also test whether rotationally-modulated noise must be considered in our analysis, and search for additional planets in the RV data set.

\subsubsection{Significance of the 7 year Periodicity}

\begin{figure*}
    \centering
    \includegraphics[width=\textwidth]{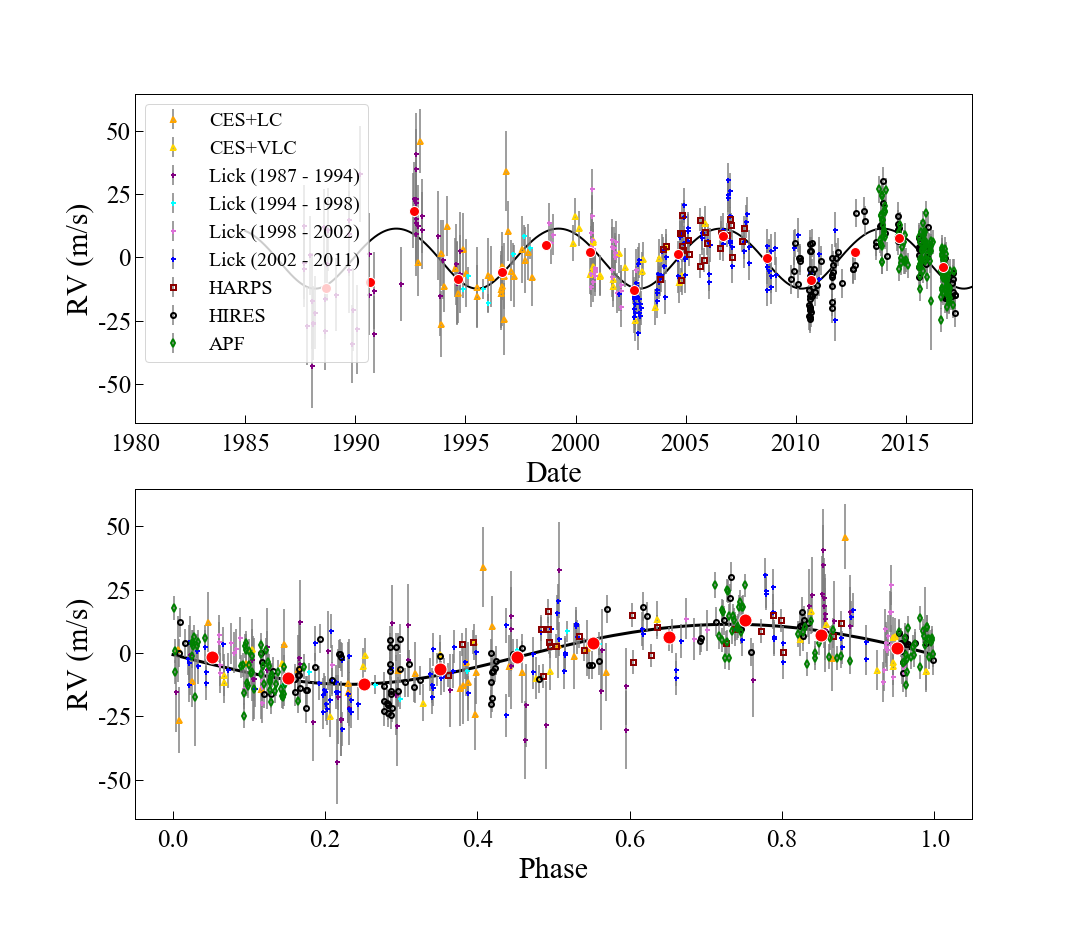}
    \caption{Time series and phase-folded radial velocity curves from all data sets are plotted. The maximum probability single-Keplerian model from \radvel\ is overplotted, as is the binned data (red). The plotted error bars include both the internal rms derived from the RV code, as well as the fitted stellar and instrumental jitter parameter $\sigma_{j}$ for each instrument.}
    \label{fig:radvel_rv}
\end{figure*}

First, we perform a 1-planet fit to the RV data using \radvel, and compare this model to the null hypothesis of no Keplerian orbit using the Bayesian Information Criterion (BIC) to determine the significance of the 7 year periodicity. The results of the \radvel\ MCMC analysis are located in Table \ref{tab:radvel}, where $P_{b}$ is the planetary orbital period, $T_{\mathrm{conj}_{b}}$ is the time of conjunction, $e_{b}$ is the planetary eccentricity, $\omega_{b}$ is the argument of periastron of the planet, and $K_{b}$ is the Keplerian semi-amplitude. $\gamma$ terms refer to the zero-point RV offset for each instrument, and $\sigma$ terms are the jitter, added in quadrature to the measurement uncertainties as described in \sectionautorefname \ref{subsec:doppler_analysis}. The maximum likelihood solution from the \radvel\ fit is plotted in Figure \ref{fig:radvel_rv} against the full RV data set. 

For the fit, the orbit is parameterized with $e_b$, $\omega_b$, $K_{b}$, $P_b$, and $T_{\mathrm{conj}_{b}}$, as well as RV offsets ($\gamma$) and jitter ($\sigma$) terms for each instrument. Due to the periodic upgrades of the Lick/Hamilton instrument and dewar, we split the \cite{Fischer2014} Lick data into 4 data sets, each with its own $\gamma$ and jitter $\sigma$ parameter. This is warranted, since \citet{Fischer2014} demonstrated that statistically significant offsets could be measured across the four upgrades in time series data on standard stars. The largest zero-point offset they measured was a 13 \ms\ offset between the third and fourth data set. Although these offsets should have been subtracted before the Lick/Hamilton data were published, the relative shifts between our derived $\gamma$ parameters match well with those reported in \citet{Fischer2014} for each upgrade, implying that the offsets were not subtracted for \thisstar.

We find that the best-fit period is $7.37\pm0.08$ years, and that this periodicity is indeed highly significant, with $\Delta \mathrm{BIC} = 245.98$ between the 1-planet model and the null hypothesis of no planets. Additionally, a model with fixed zero eccentricity is preferred ($\Delta \mathrm{BIC} = 8.5$) over one with a modeled eccentricity.

\begin{deluxetable}{lrrr}
\tablecaption{\radvel\ MCMC Posteriors}
\tablehead{\colhead{Parameter} & \colhead{Credible Interval} & \colhead{Maximum Likelihood} & \colhead{Units}}
\startdata
  $P_{b}$ & $2691^{+29}_{-28}$ & $2692$ & days \\
  $T\rm{conj}_{b}$ & $2530054^{+800}_{-770}$ & $2530054$ & JD \\
  $e_{b}$ & $0.071^{+0.061}_{-0.049}$ & $0.062$ &  \\
  $\omega_{b}$ & $3.13^{+0.82}_{-0.79}$ & $3.1$ & radians \\
  $K_{b}$ & $11.48\pm 0.66$ & $11.49$ & m s$^{-1}$ \\
\hline
  $\gamma_{\rm HIRES}$ & $2.0\pm 0.89$ & $2.05$ & \ms \\
  $\gamma_{\rm HARPS}$ & $-4.6\pm 1.6$ & $-4.8$ & \ms \\
  $\gamma_{\rm APF}$ & $-3\pm 1$ & $-3$ & \ms \\
  $\gamma_{\rm Lick_{4}}$ & $-2.45^{+0.99}_{-0.96}$ & $-2.41$ & \ms \\
  $\gamma_{\rm Lick_{3}}$ & $10.5^{+2.0}_{-1.9}$ & $11.0$ & \ms \\
  $\gamma_{\rm Lick_{2}}$ & $8.6\pm 2.3$ & $8.6$ & \ms \\
  $\gamma_{\rm Lick_{1}}$ & $8.5\pm 2.5$ & $8.5$ & \ms \\
  $\gamma_{\rm CES+LC}$ & $6.7\pm 2.7$ & $6.7$ & \ms \\
  $\gamma_{\rm CES+VLC}$ & $3.2\pm 1.8$ & $3.3$ & \ms \\
  $\dot{\gamma}$ & $\equiv0.0$ & $\equiv0.0$ & m s$^{-1}$ d$^{-1}$ \\
  $\ddot{\gamma}$ & $\equiv0.0$ & $\equiv0.0$ & m s$^{-1}$ d$^{-2}$ \\
  $\sigma_{\rm HIRES}$ & $5.99^{+0.88}_{-0.82}$ & $5.83$ & \ms \\
  $\sigma_{\rm HARPS}$ & $5.3^{+1.7}_{-1.6}$ & $4.8$ & \ms \\
  $\sigma_{\rm APF}$ & $5.26^{+0.75}_{-0.72}$ & $5.12$ & \ms \\
  $\sigma_{\rm Lick_{4}}$ & $7.1^{+0.96}_{-0.88}$ & $6.85$ & \ms \\
  $\sigma_{\rm Lick_{3}}$ & $7.6^{+1.7}_{-1.5}$ & $7.2$ & \ms \\
  $\sigma_{\rm Lick_{2}}$ & $2.8^{+2.3}_{-1.6}$ & $0.0$ & \ms \\
  $\sigma_{\rm Lick_{1}}$ & $14.5^{+2.4}_{-2.1}$ & $13.9$ & \ms \\
  $\sigma_{\rm CES+LC}$ & $9.8^{+3.0}_{-2.7}$ & $9.0$ & \ms \\
  $\sigma_{\rm CES+VLC}$ & $4.4^{+2.2}_{-2.1}$ & $3.6$ & \ms \\
\enddata
\tablenotetext{}{860000 links saved}
\tablenotetext{}{Reference epoch for $\gamma$,$\dot{\gamma}$,$\ddot{\gamma}$:       2452438.84422}
\label{tab:radvel}
\end{deluxetable}

\subsubsection{Source of the 7 year Periodicity}

We next assess whether it might be possible that the source of the periodicity at $7.37$ years is due to stellar activity, rather than a true planet.

\begin{figure*}
    \centering
    \includegraphics[width=\textwidth]{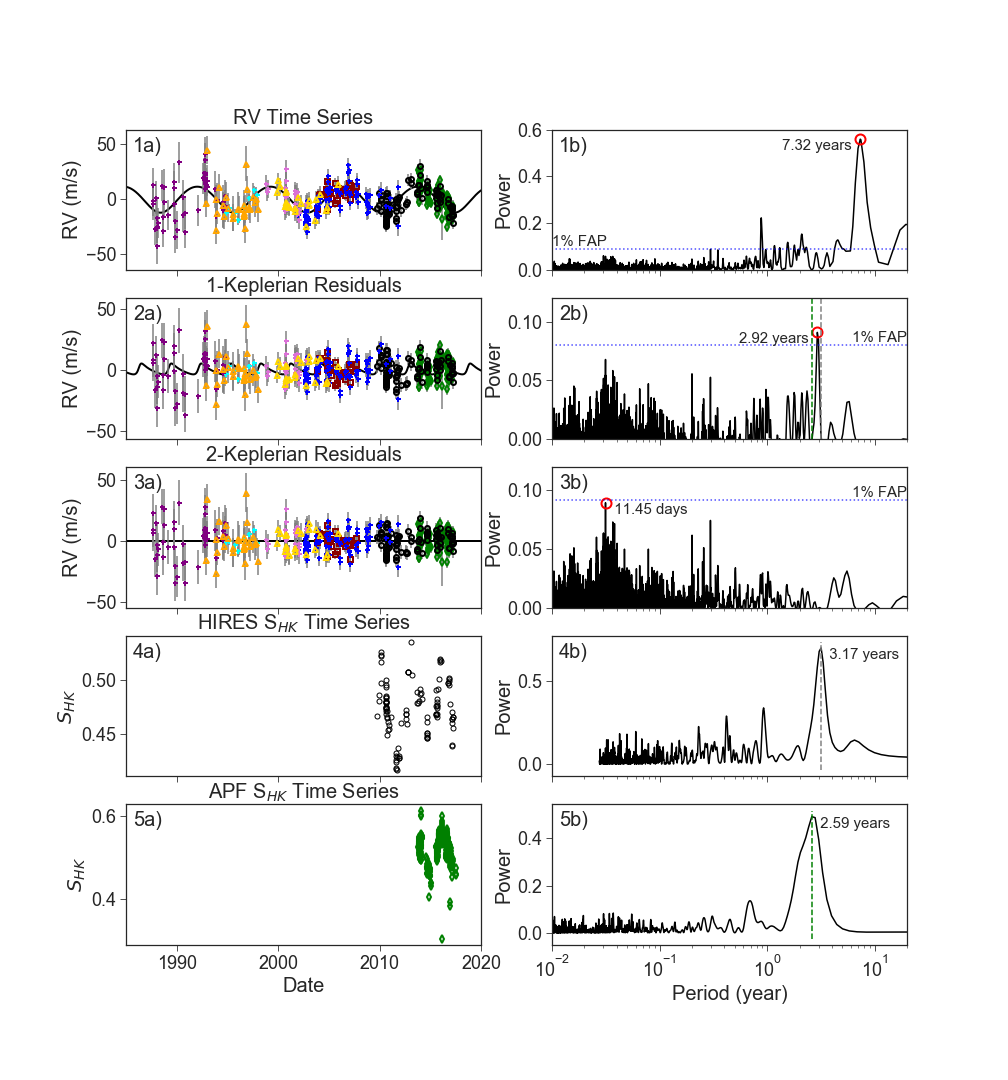}
    \caption{Time series (a) and Lomb-Scargle periodograms (b). Panels 1, 2, and 3 show the periodicities of the radial velocity measurements, residuals to a 1-Keplerian (planet) \radvel\ fit, and residuals to a 2-Keplerian (planet + stellar activity) \radvel\ fit respectively. Panels 4 and 5 show the time series and periodograms of the Keck and APF \shk values. The peak periodicities for each data set are indicated in the periodogram plots. The periodicities of the \shk data sets (panels 4--5) are overplotted in the second periodogram panel (2b), showing the correspondence between  \shk periodicity and the secondary, activity-induced peak in the RV residuals. The broad, low-significance peak at 11.45 days in panel 3b corresponds to the stellar rotation period. Plotting symbols for the RV data sets are the same as in Figure \ref{fig:radvel_rv}.}
    \label{fig:periodograms}
\end{figure*}

To probe the potential effects of the magnetic activity on the RV periodicities, we examined time series data of the RVs along with the \shk\ values from Lick, Keck, and the APF. RV and \shk\ time series and Lomb Scargle periodograms are plotted in Figure~\ref{fig:periodograms}.

A clear periodicity of 7.32 years dominates the periodogram of the RV data set. This is within the $1\sigma$ credible interval of the best-fit periodicity found with \radvel, which yielded a $\Delta \mathrm{BIC} > 200$ when we tested its significance. Once the best-fit Keplerian planetary orbit from \radvel\ is subtracted from the RV data, the residuals and their periodogram are plotted in panels 2a and 2b of Figure \ref{fig:periodograms}. The peak periodicity observed in the periodogram of the RV residuals is located at approximately 3 years, coincident the the periodicity of the \shk\ time series.

For the \shk\ time series, we detect clear \shk\ periodicities near 3 years indicative of a $\sim3$ year magnetic activity cycle (panels 4-5b). We note that the peak \shk\ periodicity appears to be slightly discrepant between the Keck and APF data sets ($P_{\rm Keck} = 3.17$ yr; $P_{\rm APF} = 2.59$ yr), but are consistent within the FWHM of the periodogram peaks. This discrepancy likely results from a variety of causes, including the shorter time baseline of the APF data, which covers only a single \shk\ cycle, and the typically non-sinusoidal and quasi-periodic nature of stellar activity cycles. The data sets also show a small offset in the median \shk\ value, likely due to differing calibrations between the instrumental and telescope setups. However, the amplitude of the \shk\ variations appears consistent between the data sets. 

We next test whether the 3 year activity cycle could be responsible for contributing power to the 7 year periodicity. The longer period is not an alias of the 3-year activity cycle, nor is it in a low-order integer ratio with the magnetic activity cycle. We perform a Keplerian fit to the RV time series from HIRES and the APF, with a period constrained at the stellar activity period (1147 days). We find an RV semi-amplitude of $K=4.8^{+2.2}_{-1.7} \ms$ and a large eccentricity of $0.53^{+0.24}_{-0.27}$ fits the data set best. We then subtract this fit from the RV data to determine whether removal of the activity-induced RV periodicity affects the significance of the planet periodicity. The 7-year periodicity in the residuals is still clearly visible by eye, and a 1-planet fit to the RV residuals after the activity cycle is subtracted yields a $\Delta \mathrm{BIC} = 197.4$ when compared to a model with no planet.

We next checked the radial velocities for correlation with \shk. Minor correlation was detected for the Keck/HIRES data set, with a Spearman correlation coefficient of $r_S = 0.28$ at moderate statistical significance ($p = 0.01$). For the APF data set, a stronger and statistically significant correlation was found ($r_S=0.50$, $p<<0.01$) between \shk\ and RV. However, given that the 3-year magnetic cycle shows up in the radial velocity residuals, it is not surprising that RV and \shk\ might be correlated. There is also rotationally-modulated noise that might be present in both data sets near the $\sim$ 11-day rotation timescale, increasing the correlation. We attempt to determine whether the measured correlation derives from the 3-year periodicity in both data sets, or whether it is produced by other equivalent periodicities in the RV and \shk\ data sets.

To test this, we first performed a Keplerian fit with a period of approximately 3 years to the Keck and APF \shk\ time series using \radvel. Although stellar activity is not the same as orbital motion, we used the Keplerian function as a proxy for the long-term stellar activity cycle of \thisstar. We found a maximum probability period of $1194^{+30}_{-25}$ days for the HIRES data and $989^{+40}_{-26}$ days for the APF data. These periodicities are indeed discrepant by more than $5\sigma$. When combined, we found a period of $1147^{+22}_{-20}$ days for the full HIRES and APF data set. 

We then subtracted the maximum probability 3-year fit from each \shk\ data set, and examined the residual values. We found that the correlation between these \shk\ residuals and the radial velocity data was significantly reduced for the APF data, with $r_S=0.17$ and $p=0.04$. This suggests that the strong correlation we detected was primarily a result of the 3-year periodicity. For the HIRES data set, the moderately significant correlation of $r_S = 0.28$ was unchanged. 

\begin{figure}
    \centering
    \includegraphics[width=0.5\textwidth]{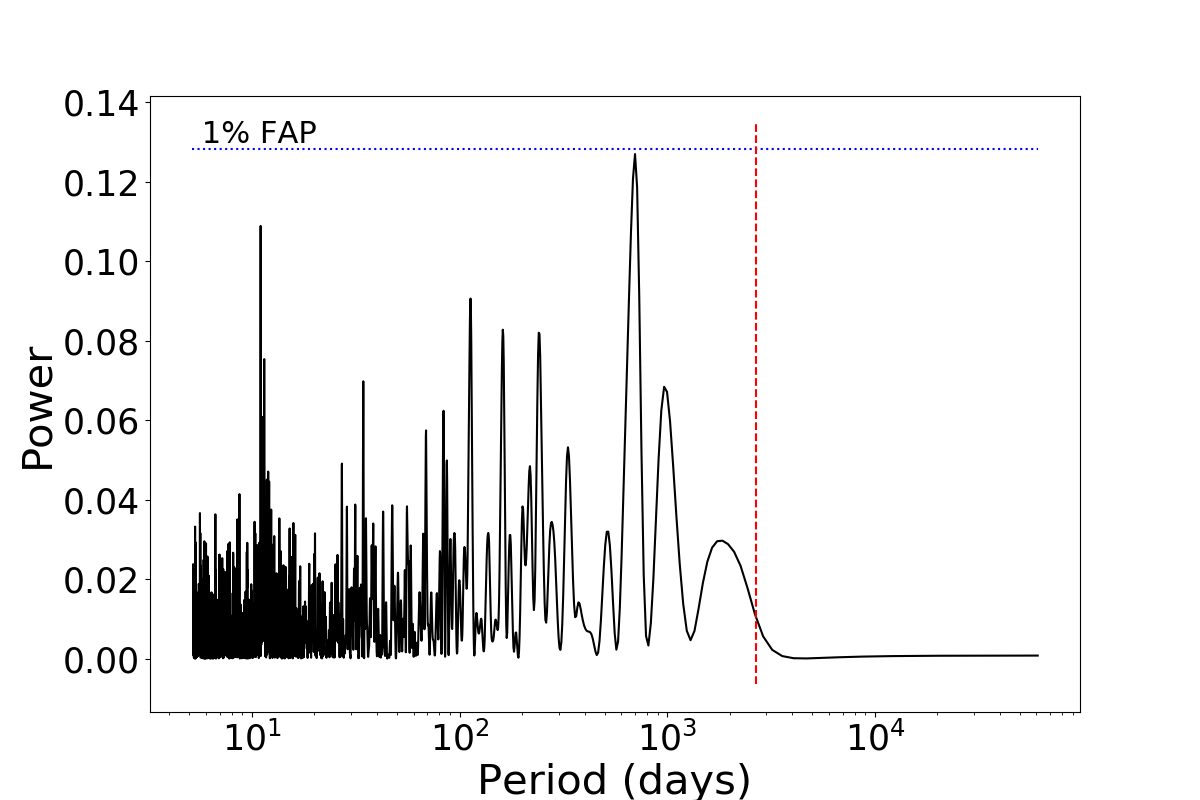}
    \caption{Periodogram of the HIRES and APF \shk\ residuals to the $\sim 3$ year fit. The red dotted line shows the best-fit period of the planet from our initial 1-planet fit. Like the \shk\ periodograms shown in Figure \ref{fig:periodograms} panels 4-5b, there is no power at the planet's orbital period. Even when the peak periodicity is removed for each data set, no additional power appears at the planet's 7.37 year orbital period. This indicates that stellar activity is not likely to cause the 7.37 year periodicity in the radial velocity data.}
    \label{fig:sresid_peri}
\end{figure}

A Lomb-Scargle periodogram of the \shk\ residuals is displayed in Figure \ref{fig:sresid_peri}. It shows no significant peak or power near the posterior planet period at 2691 days, demonstrating that the \shk\ time series has no significant periodicity at the planet's orbital period.

Our next tests involved modifying our 1-planet fits to the RV data to account for the stellar activity cycle in two ways. We first performed a 1-planet fit to the RVs using a linear decorrelation against the \shk\ values for the HIRES and APF data sets. We then performed a 2-Keplerian fit to the RV data, in order to simultaneously characterize both the planetary orbit and the stellar activity cycle. In both cases, we checked for significant changes to the planetary orbital parameters due to accounting for the stellar activity cycle in the fit. For both tests, we find that the maximum likelihood values of the planet's orbital parameters all agree within $1\sigma$ credible intervals with the single-planet fit.

We note that from the 2-Keplerian fit, the best-fit second Keplerian provides some information about the stellar activity cycle. It has a best-fit period of $1079$ days, or $2.95$ years, shorter than the periodicity derived from a fit to the HIRES and APF \shk\ time series. However, though the other parameters in this fit seem to be converged, the period and time of conjunction of the second Keplerian are clearly not converged over the iterations completed for this model. Increasing the number of iterations does not appear to improve convergence. This again points to the quasi-periodic nature of stellar activity cycles, and the different time baselines of the full RV data set and the \shk\ time series available. The fit has an RV semi-amplitude of $K_{activity} = 4.4$ \ms, lower than the semi-amplitude of the planet at $K_{b} = 11.81 \pm 0.65$ \ms.

\subsubsection{Search for additional RV planets}

We use the automated planet search algorithm described by \cite{Howard2016} to determine whether additional planet signatures are present in the combined radial velocity dataset. The residuals to the 2-Keplerian fit are examined for additional periodic signatures. The search is performed using a 2D Keplerian Lomb-Scargle periodogram \citep[2DKLS,][]{OToole2009}.

The residuals to the 2-Keplerian fit show several small peaks, but none with empirical false alarm probabilities \citep[eFAP,][]{Howard2016} less than 1\% (Figure~\ref{fig:periodograms}, panel 3b). A broad forest of peaks at approximately 12 days corresponds to the stellar rotation period, and is likely due to spot-modulated stellar jitter. The next most significant peak is located at 108.3 days. We attempt a 3-Keplerian fit to the RVs with the third Keplerian initiated at 108.3 days. However, we are unable to achieve convergence in a reasonable number of iterations, and the walkers are poorly-behaved. This serves as evidence against the inclusion of a third periodicity. We conclude that there is insufficient evidence to suggest an inner planet to \thisstar\ b exists.

RV and residual time series, as well as 2DKLS periodograms used for the additional planet search, are plotted in Figure \ref{fig:periodograms}, panel 3a and 3b.

\subsubsection{Gaussian Processes Fits}

For all of these analyses, we have assumed white noise and added a ``jitter'' term in quadrature to account for uncertainty due to stellar activity. We assessed whether this was reasonable by performing a 1-planet fit using \radvel\ and including a Gaussian processes model to account for rotationally-modulated stellar noise \citep{Blunt2018,Kosiarek2018} as well as the 3 year stellar activity cycle.

First we used \radvel\ with a new implementation of GP regression using the quasi-periodic covariance kernel to fit four GP hyperparameters in addition to the Keplerian parameters for a single planet and a white noise term $\sigma_{j}$. The hyperparameters for the quasi-periodic kernel are the amplitude of the covariance function ($h$); the period of the correlated noise ($\theta$, in this case trained on the rotation period of the star); the characteristic decay timescale of the correlation ($\lambda$, a proxy for the typical spot lifetime); and the coherence scale ($w$, sometimes called the structure parameter) \citep{Grunblatt2015,Lopez-Morales2016}.

We applied a Gaussian prior to the rotation period of $\theta = 11.45\pm2.0$ days, based on the periodicity observed in the radial velocity residuals to the 2-Keplerian fit, but sufficiently wide to allow the model flexibility. The covariance amplitudes $h$ for each instrument were constrained with a Jeffrey's prior truncated at 0.1 and 100 \ms. We imposed a uniform prior of 0--1 years on the exponential decay timescale parameter $\lambda$. We chose a Gaussian prior for $w$ of $0.5\pm 0.05$, following \citet{Lopez-Morales2016}.

The results of our GP analysis provide constraints on the hyperparameters, indicating that the rotation period is $11.64^{+0.33}_{-0.24}$ days and the exponential decay timescale is $49^{+15}_{-11}$ days. The amplitude parameters for each instrument ranged from 0.0 \ms\ to 13.4 \ms, and were highest for the earliest Lick RV data. For some of the data sets, the cadence of the observations likely reduced their sensitivity to correlated noise on the rotation timescale, resulting in GP amplitudes consistent with zero. For other instruments, notably the HIRES and APF data, the white noise jitter term $\sigma_{j}$ was significantly reduced in the GP model, compared with the standard RV solution. 

However, when comparing the derived properties of the planet, we find that the GP analysis has no noticeable effect on the planet's  orbital parameters. The period, RV semi-amplitude, eccentricity, time of conjunction, and argument of periastron constraints from the GP regression analysis all agree within $1\sigma$ with the values derived from the traditional 1-planet fit. We therefore conclude that the rotationally-modulated noise does not significantly affect the planet's orbital parameters. 

We additionally performed a 1-planet fit using GP regression to model the 3-year stellar activity cycle. For this test case, we used a periodic GP kernel, since each data set covers only a relatively few cycles of the stellar activity cycle. Unlike activity signatures at the stellar rotation period, we do not expect to see significant decay or decorrelation of the 3 year cycle over the time span of our data set. This periodic GP model had hyperparameters describing the periodicity ($\theta$), amplitude ($h$), and structure parameter ($w$), but no exponential decay. This analysis is somewhat akin to our 2-Keplerian fit, but allows more flexibility to fit the noise than a Keplerian. For this model, we placed a Gaussian prior of $\theta = 1147\pm20$ days on the GP period parameter, based on the Keplerian fit to the \shk\ values. We found that when allowing each instrument its own GP amplitude parameter, $h$, nearly all of the instrumental amplitudes were best fit with values very close to zero, so instead we fit for only a single GP amplitude across all instrumental data sets. We constrained this parameter with a Jeffreys prior bounded at 0.01 -- 100 \ms. We again used the $w=0.5\pm0.05$ prior for the structure parameter, after testing out fits at several values between 0 and 1. It remains unclear whether this was the optimal choice, since the physical interpretation of this parameter would be different for the long-term stellar activity cycle as compared with the rotationally-modulated spot noise.

The results of this analysis indicate a GP periodicity of $\theta = 1149\pm17$ days, a slightly tighter constraint than the imposed prior. The GP amplitude parameter was constrained to be $h = 4.26 ^{+1.21}_{-1.01}$ \ms, comparable with the posteriors for the RV semi-amplitude of the second Keplerian in the 2-Keplerian fit. Importantly, the model posteriors on the planetary parameters were again consistent within $1\sigma$ with our traditional 1-planet fit for all parameters, including the white noise jitter terms $\sigma_{j}$ as well as the orbital period, RV semi-amplitude, eccentricity and Keplerian angles. 

We note that the traditional 1-planet fit is preferred over the 1-planet Gaussian processes fit by $\Delta \mathrm{BIC} = 19.7$. The 2-Keplerian fit is also preferred over the GP 1-planet model, with $\Delta \mathrm{BIC} = 30.6$, despite having 2 additional free parameters and nominally less flexibility than the GP model.

These tests demonstrate that the addition of a Keplerian or Gaussian Processes model to account for stellar activity (both rotationally-modulated activity and the long-period stellar activity cycle) does not strongly influence the results of the planetary orbital fit. The GP fit in particular was statistically disfavored compared to the simpler Keplerian model based on $\Delta$ BIC. We therefore choose to restrict our subsequent analyses to consider only a single planet, and only white noise. Going forward, the uncertainty due to stellar activity is added in quadrature as a white-noise ``jitter'' term, and red noise is not considered.


\subsection{Combining constraints from imaging and RV} \label{subsec:results3}

By combining the imaging and radial velocity data sets, it is possible to place tighter upper limits on the mass of the companion. Indeed, the RV data provides a lower limit on the planet mass ($M\sin{i}$), while the direct imaging data complements it with an upper limit.

A MCMC will be used to infer the posterior on the masses and orbital parameters of the system, noted $\Theta$. The noise in the radial velocity measurements $d_{RV}$ and in the images $d_{DI}$ is independent, which means that the joint likelihood is separable such as
\begin{multline}
   \mathcal{P}(d_{DI},d_{RV}\vert \Theta) = \mathcal{P}(d_{RV}\vert \Theta)\mathcal{P}(d_{DI}\vert \Theta)
  .
    \label{eq:joint_likelihood}
\end{multline}

\subsubsection{Direct imaging likelihood} \label{subsec:di_like}

In this section, we detail the computation of the direct imaging likelihood \citep{Ruffio2018}. 
The direct imaging data $d_{DI}$, temporarily shorten $d$, is a vector of $N_{exp}\times N_{pix}$ elements where $N_{exp}$ is the number of exposures in the dataset and $N_{pix}$ the number of pixels in an image. It is the concatenation of all the vectorized speckle subtracted single exposures.
A point source is defined from its position $x$ and its brightness $i$. We also define $n$ as a Gaussian random vector with zero mean and covariance matrix $\Sigma$. We assume that the noise is uncorrelated and that $\Sigma$ is therefore diagonal.
\begin{equation}
d=im+n
\end{equation}
With $m=m(x)$ being a normalized planet model at the position $x$.

Assuming Gaussian noise, the direct imaging likelihood is given by:
\begin{align}
\mathcal{P}(d\vert i, x) &= \frac{1}{\sqrt{2\pi\vert\bm{\Sigma}\vert}} \exp\left\lbrace-\frac{1}{2}(\bm{d}-i\bm{m})^{\top}\bm{\Sigma}^{-1}(\bm{d}-i\bm{m})\right\rbrace \nonumber \\
&\propto
\exp\left\lbrace-\frac{1}{2}\left(i^2\bm{m}^{\top}\bm{\Sigma}^{-1}\bm{m}-2i\bm{d}^{\top}\bm{\Sigma}^{-1}\bm{m}\right)\right\rbrace 
\label{eq:likelihood}
\end{align}
We have used the fact that $\bm{d}^{\top}\bm{\Sigma}^{-1}\bm{d}$ is a constant, because we are not inferring the direct imaging covariance.

The estimated brightness $\tilde{i}_x$, in a maximum likelihood sense, and associated error bar $\sigma_x$ are defined as,
\begin{equation}
\tilde{i}_x = \frac{\bm{d}^{\top} \bm{\Sigma}^{-1} \bm{m} }{\bm{m}^{\top} \bm{\Sigma}^{-1} \bm{m}},
\label{eq:MLEa}
\end{equation}
and
\begin{equation}
\sigma_{x}^2 = (\bm{m}^{\top} \bm{\Sigma}^{-1} \bm{m})^{-1}.
\end{equation}
We can therefore rewrite the logarithm of the direct imaging likelihood as a function of these quantities \citep{Ruffio2018},
\begin{equation}
    \log{\mathcal{P}(d\vert i, x)} = -\frac{1}{2\sigma_x^2}\left(i^2-2i\tilde{i}_x\right).
    \label{eq:dilikelihood}
\end{equation}

The definition of the planet model $m$ is challenging when using a PCA-based image processing. Indeed, while it subtracts the speckle pattern, it also distorts the signal of the planet. 
The distortion is generally not accounted for in a classical data reduction such as the one used in \sectionautorefname \ref{subsec:results1}, which is why it is more convenient to adopt a Forward Model Matched Filter (FMMF) approach as described in \citet{Ruffio2017}. The FMMF computes the map of estimated brightness and standard deviation used in \autoref{eq:dilikelihood} by deriving a linear approximation of the distorted planet signal for each independent exposure, called the forward model \citep{Pueyo2016}.

We showed that the likelihood can theoretically be calculated directly from the final products of the FMMF. In practice, the noise is correlated and not perfectly Gaussian resulting in the standard deviation to be underestimated and possibly biasing the estimated brightness. We therefore recalibrate the S/N by dividing it by its standard deviation computed in concentric annuli. The estimated brightness map is corrected for algorithm throughput using simulated planet injection and recovery. The likelihood is computed for the fully calibrated SNR maps.

FMMF is part of a Python implementation of the PCA algorithm presented in \citet{Soummer2012} called \PyKLIP \footnote{Available under open-source license at \url{https://bitbucket.org/pyKLIP/pyklip}.} \citep{Wang2015}.
The principal components for each exposure are calculated from a reference library of the 200 most correlated images from which only the first 20 modes are kept. Images in which the planet would be overlapping with the current exposure are not considered to be part of the reference library to limit the self- and over-subtraction using an exclusion criterion of $7$ pixels ($0.7\lambda/D$). The speckle subtraction is independently performed on small sectors of the image.

\subsubsection{Joint likelihood and priors} \label{subsec:joint_like} 

We implement a Markov-chain Monte Carlo analysis of the combined radial velocity data from the Coud\'e Echelle Spectrograph, HARPS, Lick/Hamilton, Keck/HIRES, and  APF/Levy instruments, as well as the single-epoch direct imaging data. We solve for the full Keplerian orbital parameters, including orbital inclination and longitude of the ascending node, which are not typically included in RV-only orbital analyses. Including the full Keplerian parameters allows us to calculate the projected position of the companion at the imaging epoch for each model orbit. This is necessary to calculate an additional likelihood based on the direct imaging data. 

The full log-likelihood function used for this analysis is:

\begin{multline}
   \log{\mathcal{P}(d_{DI},d_{RV}\vert \Theta)} = - \frac{1}{2\sigma_x^2}\left(i^2-2i\tilde{i}_x\right) \\ -\sum_i{\left[\frac{(v_i-v_m(t_i))^2}{2(\sigma_i^2 + \sigma_j^2)} + \log{\sqrt{2\pi(\sigma_i^2+\sigma_j^2)}}\right]}
  .
    \label{eq:full_likelihood}
\end{multline}

The RV component of the likelihood comes from  \citep{Howard2014}. Here, $v_i = v_{i,inst} - \gamma_{inst}$ is the offset-subtracted radial velocity measurement, and $\sigma_i$ refers to the internal uncertainty for each measurement. $v_m(t_i)$ is the Keplerian model velocity at the time of each observation; $\sigma_j$ is the instrument-specific jitter term, which contributes additional uncertainty due to both stellar activity and instrumental noise. In these models, each instrument's radial velocity offset ($\gamma_{inst}$) and jitter term ($\sigma_{j,inst}$) are included as free parameters in the fit. A description of the direct imaging component of the likelihood is available in \sectionautorefname \ref{subsec:di_like}.

We draw from uniform distributions in $\log{P}$, $\log{M_{b}}$, $\cos{i}$, $\sqrt{e}\cos{\omega}$, $\sqrt{e}\sin{\omega}$, $\Omega$, mean anomaly at the epoch of the first observation, and $\gamma_{inst}$. 

We place a tight gaussian prior of $M_{\star} = 0.781 \pm 0.078\ M_{\odot}$ on the primary stellar mass, based on the interferometric results of \citet{Boyajian2012}. Other groups have measured slightly different but generally consistent stellar masses for \thisstar. \citet{Valenti2005} report a spectroscopic mass of $M_{\star} = 0.708 \pm 0.067 M_{\odot}$; \citet{Takeda2007} report a discrepant spectroscopic result of $M_{\star} = 0.856^{+0.06}_{-0.08} M_{\odot}$.

A tight gaussian prior of $\pi = 310.94 \pm 0.16$ mas is also imposed on stellar parallax based on the Hipparcos parallax measurement for this star \citep{vanLeeuwen2007}. We place wide gaussian priors on the jitter terms, with $\sigma_{j} = 10.0 \pm 10.0\ \ms$. Large values for jitter are also disfavored by the second term of the likelihood function.

With these priors and likelihood function, we solve for the full orbital parameters and uncertainties using the python package \emcee\ \citep{Foreman-Mackey2013}. For comparison with the \radvel\ results, we perform our analysis both with and without the direct imaging likelihood. We use planet models of ages 800 Myr, 400 Myr, and 200 Myr in individual analyses, since the system's age constraints span this range. We use the standard \emcee\ Ensemble Sampler; each MCMC run uses 100 walkers, and is iterated for more than 500000 steps per walker. We check that each sampler satisfies a threshold of Gelman-Rubin statistic $\hat{R} < 1.1$ for all parameters \citep{Ford2006,Gelman1992} to test for nonconvergence. We note that average acceptance fractions for our chains are fairly low, $\approx5 - 10$\%.

\subsubsection{MCMC Results}
\label{mcmc_results}
\begin{figure}
    \centering
     \includegraphics[width=0.5\textwidth]{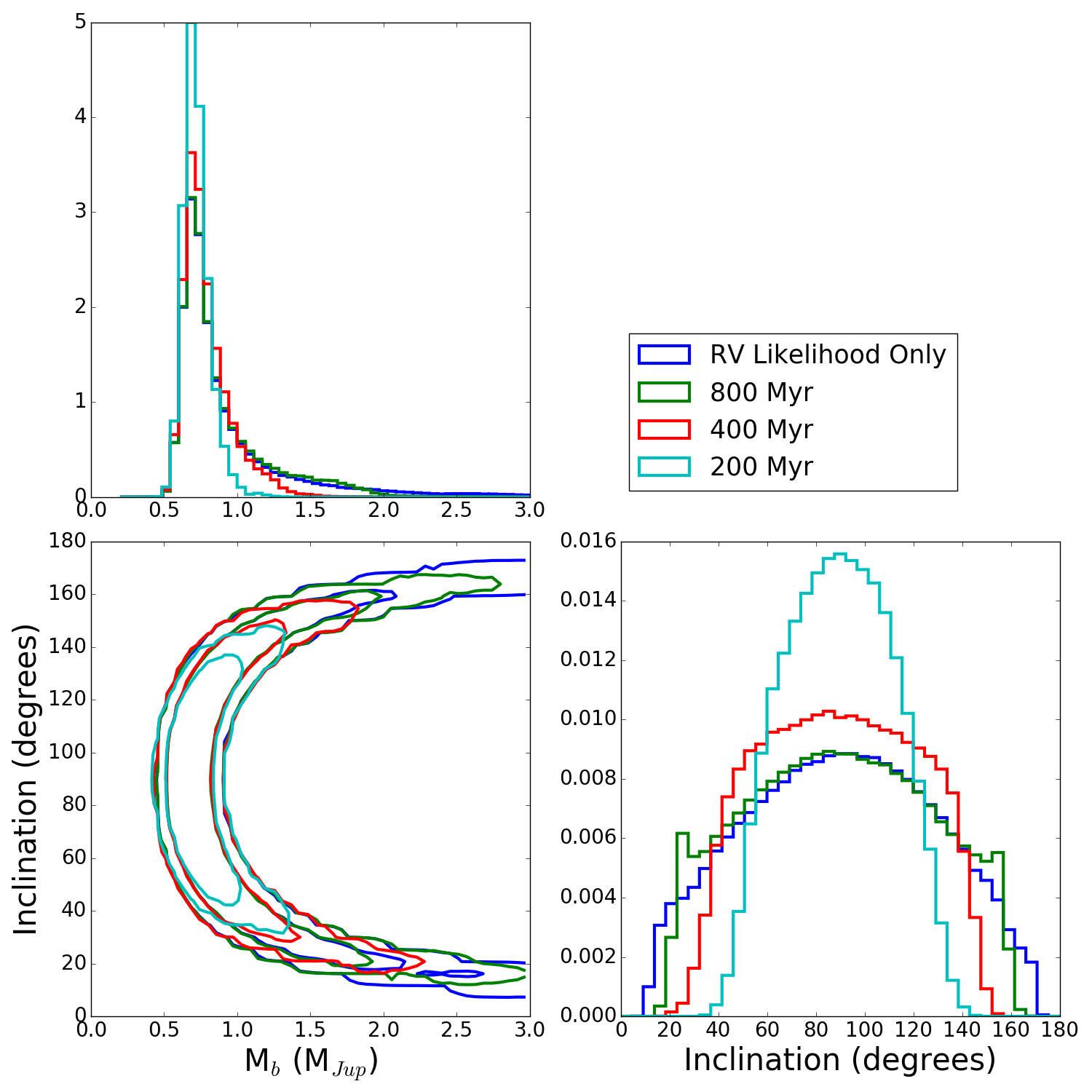}
     \caption{Corner plot showing the posterior distributions and correlation between the companion mass and inclination for models using the RV likelihood only, as well as RV + direct imaging likelihood with planet models of age 800, 400, and 200 Myr (a log-uniform prior).}
     \label{fig:m2_inc_corner}
 \end{figure}

The planet parameters derived in this analysis are consistent with those determined by \radvel. The posterior distributions for the companion mass and orbital inclination are plotted in Figure \ref{fig:m2_inc_corner}. The lower limit on planet mass $M_{b}\sin{i}=0.72\pm{0.07}$ $M_{Jup}$ is constrained by the Keplerian velocity semi-amplitude and agrees well with the \radvel\ results. With the RV data alone, the true mass (independent of $\sin{i}$) has a poorly constrained upper limit, although high-mass, low-inclination orbits are geometrically disfavored. With the addition of the imaging nondetection constraints, the mass upper limit is improved.

Since younger planets are hotter and thus brighter, the direct imaging likelihood disfavors a broader region of parameter space when a younger age is assumed. Thus, the tightest constraints come from the youngest-aged planet models. Table \ref{tab:mcmc_res} lists the planet parameters resulting from each MCMC run. We report the median and 68\% credible intervals for each model.

We also calculate the posterior distribution on the position of the planet at the epoch of the NIRC2 imaging observation from the RV-only likelihood model. We check this posterior to ensure that the imaging observations were optimally timed to detect the planet at maximal separation from the star. The positional posterior distribution is plotted in Figure \ref{fig:xy_posterior}, and demonstrates that at the epoch of the imaging observations, the separation of the planet from the star was indeed maximized. The planet would have been easily resolvable, regardless of the on-sky orientation (longitude of the ascending node).

For these analyses, we draw companion mass uniformly in logarithmic space with bounds at 0.01 and 100 $M_{Jup}$. This is comparable to placing a Jeffreys prior, a common choice of prior for scale parameters such as mass and period \citep{Ford2006}. This prior is also not significantly dissimilar to the mass distribution of Doppler-detected Jovian planets from  \citet{Cumming2008}, who found that $\frac{dN}{d\log{m}} \propto M^{-0.31}$, a roughly flat distribution in $\log{m}$.

\begin{figure}
    \centering
     \includegraphics[width=0.5\textwidth]{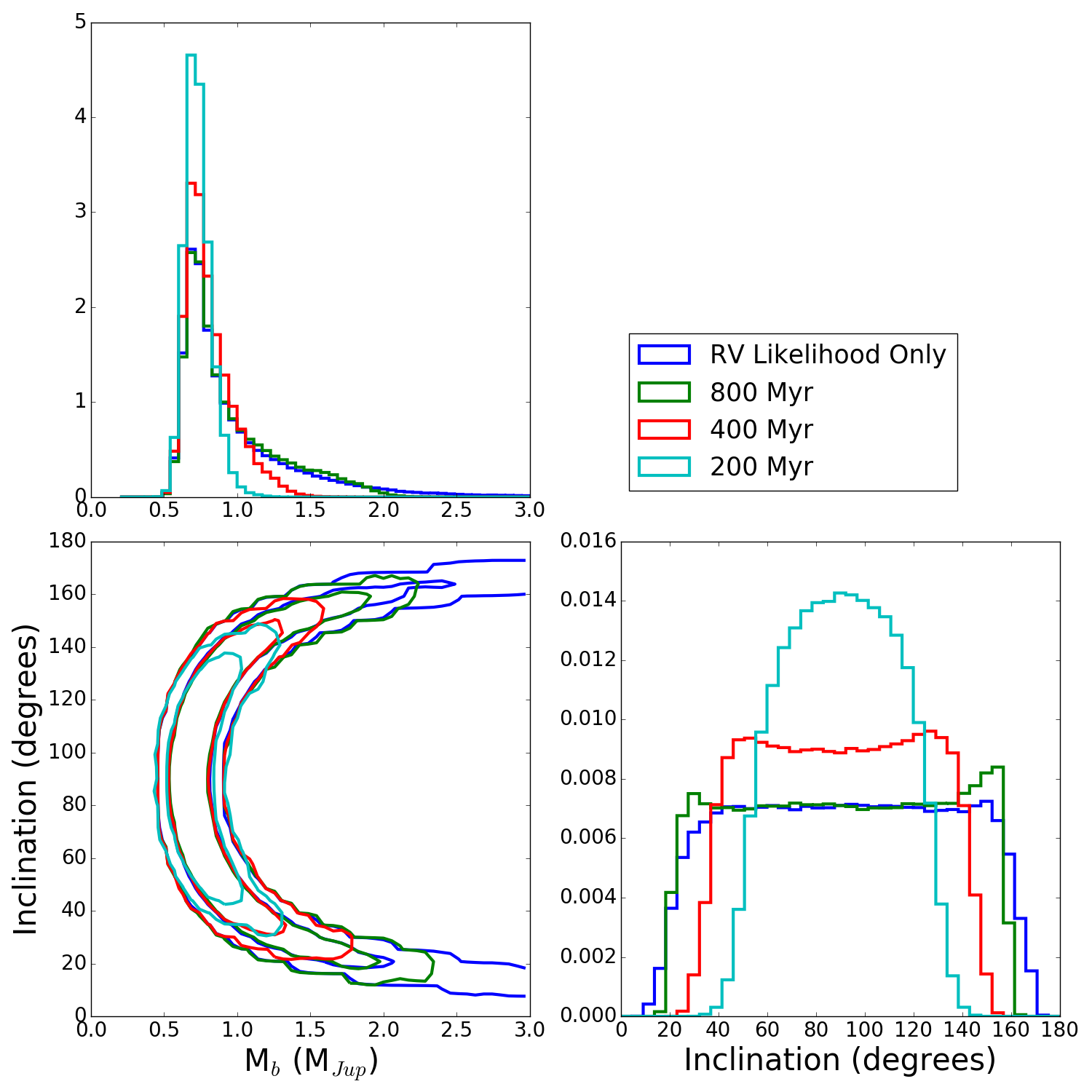}
     \caption{Same as Figure \ref{fig:m2_inc_corner}, but with uniform prior on companion mass.}
     \label{fig:m2_inc_corner_lin}
 \end{figure}

\begin{figure}
    \centering
    \includegraphics[width=0.5\textwidth]{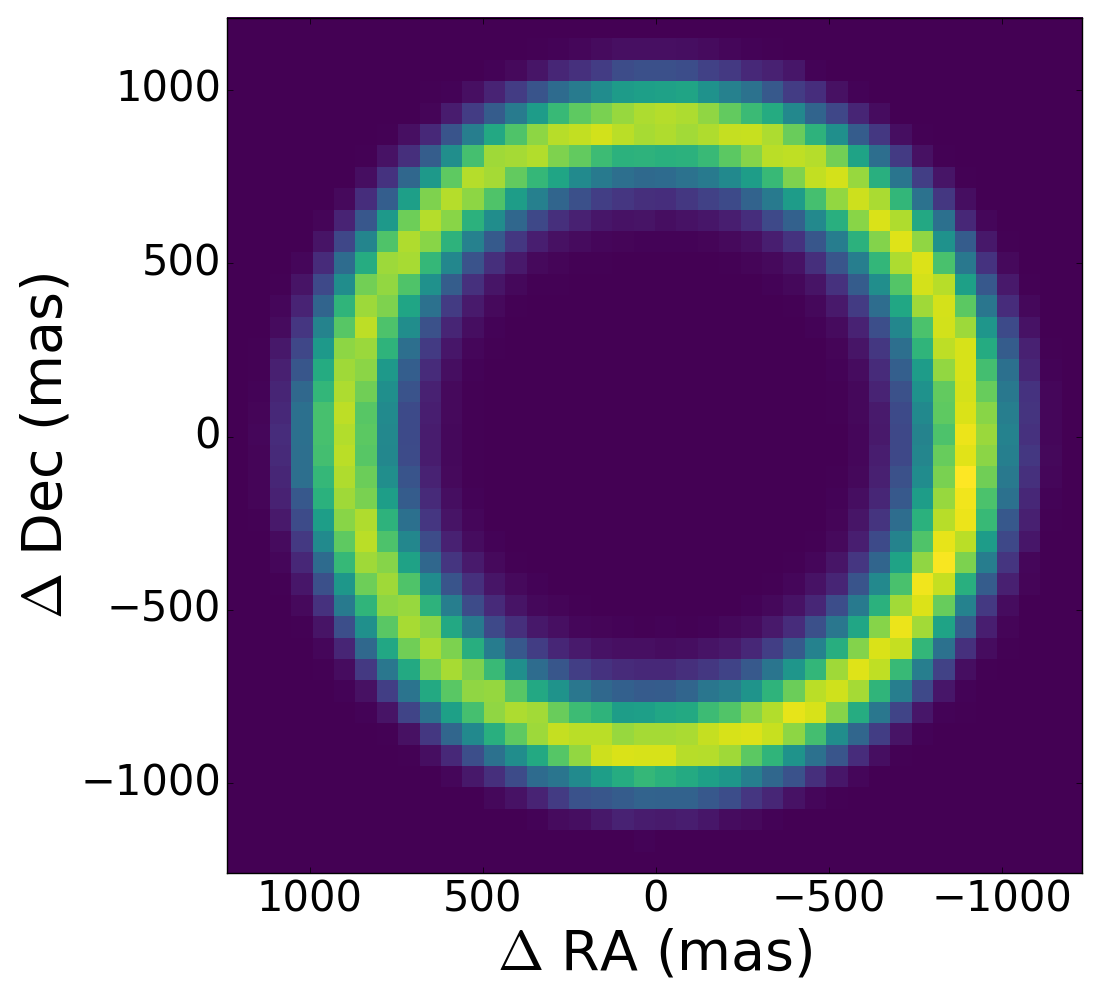}
    \caption{2-Dimensional posterior distribution of the position of the planet during the epoch of the imaging observations. This posterior was produced using the RV likelihood only, and demonstrates that the planet was optimally separated from its host star at the imaging epoch. The values of the pixels in the maximum-likelihood annulus contribute most significantly to the direct imaging likelihood.}
    \label{fig:xy_posterior}
\end{figure}

To assess the impact of this choice, we repeat our analysis with a uniform prior on the mass, again from 0.01 to 100 $M_{Jup}$. This alternative increases the significance of the tail of the $m_{b}$ posterior distribution toward higher masses. Since mass and inclination are highly correlated, this effect also serves to flatten out the inclination posterior, adding more significance to lower-inclination orbits. Figure \ref{fig:m2_inc_corner_lin} shows the posteriors and correlation between the mass and inclination of the planet under the modified mass prior. The correlation plot is identical to that shown in Figure~\ref{fig:m2_inc_corner}, and the mass posterior is not qualitatively changed. The median / 68\% confidence interval planet mass from the 800 Myr model is m$_{b} = 0.83^{+0.47}_{-0.15} M_{Jup}$, consistent within uncertainties with the mass constraint from the log-mass case at the same age. The inclination posterior has a wider uncertainty in the linear mass case ($i = 90.8^{\circ}\pm48.0 ^{\circ}$, as compared with the log mass case ($i = 89.2^{\circ} \pm 41.7 ^{\circ}$). All other orbital and instrumental parameters have equivalent constraints in both cases. We conclude that the prior on mass does not significantly affect the results of the analysis.

\begin{table*}[t]
\centering
\footnotesize
\caption{MCMC Results} 
\label{tab:mcmc_res}
\begin{tabular}{lcccc}
\hline
& RV Likelihood Only & 800 Myr & 400 Myr & 200 Myr \\
\hline
Parameter & \multicolumn{4}{c}{Median and 68\% Credible Interval} \\
\hline
m$_{b}$ (M$_{Jup}$) & 
$0.78^{+0.43}_{-0.12}$ & 
$0.78^{+0.38}_{-0.12}$ & 
$0.75^{+0.19}_{-0.10}$ & 
$0.71^{+0.09}_{-0.07}$ \\

P (yr) & 
$7.37^{+0.07}_{-0.07}$ & 
$7.37^{+0.07}_{-0.07}$ & 
$7.37^{+0.07}_{-0.07}$ & 
$7.38^{+0.07}_{-0.07}$ \\

e & 
$0.07^{+0.06}_{-0.05}$ & 
$0.07^{+0.06}_{-0.05}$ & 
$0.07^{+0.06}_{-0.05}$ & 
$0.06^{+0.06}_{-0.04}$ \\

$\omega$ ($^{\circ}$) & 
$177^{+49}_{-51}$ & 
$175^{+53}_{-52}$ & 
$177^{+48}_{-49}$ & 
$157^{+66}_{-51}$ \\

$\Omega$ ($^{\circ}$) & 
$180^{+122}_{-123}$ & 
$184^{+126}_{-131}$ & 
$212^{+108}_{-148}$ & 
$276^{+47}_{-158}$ \\

i ($^{\circ}$) & 
$90^{+42}_{-43}$ & 
$89^{+42}_{-42}$ & 
$89^{+35}_{-35}$ & 
$90^{+23}_{-24}$ \\

t$_{peri}$ (JD) & 
$2447213^{+336}_{-429}$ & 
$2447198^{+361}_{-426}$ & 
$2447218^{+332}_{-407}$ & 
$2447032^{+475}_{-402}$ \\
\hline

$\gamma_{Lick 1}$ (\ms) & 
$8.4^{+2.4}_{-2.4}$ & 
$8.5^{+2.4}_{-2.4}$ & 
$8.4^{+2.4}_{-2.3}$ & 
$8.4^{+2.4}_{-2.4}$ \\

$\sigma_{Lick 1}$ (\ms) & 
$15.1^{+2.0}_{-1.8}$ & 
$15.1^{+2.0}_{-1.8}$ & 
$15.1^{+2.1}_{-1.8}$ & 
$15.1^{+2.1}_{-1.8}$ \\

$\gamma_{CES+LC}$ (\ms) & 
$6.9^{+2.6}_{-2.6}$ & 
$6.9^{+2.6}_{-2.6}$ & 
$6.9^{+2.6}_{-2.6}$ & 
$6.9^{+2.6}_{-2.5}$ \\

$\sigma_{CES+LC}$ (\ms) & 
$10.9^{+2.5}_{-2.1}$ & 
$10.9^{+2.5}_{-2.1}$ & 
$10.9^{+2.5}_{-2.2}$ & 
$10.9^{+2.5}_{-2.2}$ \\

$\gamma_{Lick 2}$ (\ms) & 
$8.5^{+1.9}_{-1.9}$ & 
$8.5^{+1.9}_{-1.9}$ & 
$8.5^{+1.9}_{-1.9}$ & 
$8.5^{+2.0}_{-1.9}$ \\

$\sigma_{Lick 2}$ (\ms)  & 
$4.7^{+2.1}_{-1.3}$ & 
$4.7^{+2.2}_{-1.4}$ & 
$4.7^{+2.1}_{-1.4}$ & 
$4.8^{+2.2}_{-1.4}$ \\

$\gamma_{Lick 3}$ (\ms) & 
$10.5^{+1.9}_{-1.9}$ & 
$10.6^{+1.8}_{-1.9}$ & 
$10.5^{+1.9}_{-1.9}$ & 
$10.6^{+1.9}_{-1.8}$ \\

$\sigma_{Lick 3}$ (\ms) & 
$9.2^{+1.4}_{-1.2}$ & 
$9.2^{+1.4}_{-1.1}$ & 
$9.2^{+1.4}_{-1.1}$ & 
$9.2^{+1.4}_{-1.1}$ \\

$\gamma_{CES+VLC}$ (\ms) & 
$3.4^{+1.7}_{-1.8}$ & 
$3.4^{+1.8}_{-1.9}$ & 
$3.4^{+1.8}_{-1.8}$ & 
$3.4^{+1.8}_{-1.8}$ \\

$\sigma_{CES+VLC}$ (\ms) & 
$6.8^{+1.8}_{-1.5}$ & 
$6.8^{+1.8}_{-1.5}$ & 
$6.9^{+1.7}_{-1.5}$ & 
$6.8^{+1.8}_{-1.5}$ \\

$\gamma_{Lick 4}$ (\ms) & 
$-2.4^{+1.0}_{-1.0}$ & 
$-2.4^{+1.0}_{-1.0}$ & 
$-2.4^{+1.0}_{-1.0}$ & 
$-2.4^{+1.0}_{-1.0}$ \\

$\sigma_{Lick 4}$ (\ms) & 
$8.7^{+0.8}_{-0.7}$ & 
$8.7^{+0.7}_{-0.7}$ & 
$8.7^{+0.7}_{-0.7}$ & 
$8.7^{+0.7}_{-0.7}$ \\

$\gamma_{HARPS}$ (\ms) & 
$-4.5^{+1.6}_{-1.6}$ & 
$-4.5^{+1.5}_{-1.5}$ & 
$-4.5^{+1.5}_{-1.5}$ & 
$-4.4^{+1.6}_{-1.6}$ \\

$\sigma_{HARPS}$ (\ms) & 
$7.4^{+1.3}_{-1.0}$ & 
$7.4^{+1.3}_{-1.0}$ & 
$7.4^{+1.3}_{-1.0}$ & 
$7.4^{+1.3}_{-1.0}$ \\

$\gamma_{HIRES}$ (\ms) & 
$2.0^{+0.9}_{-0.9}$ & 
$2.0^{+0.9}_{-0.8}$ & 
$2.0^{+0.9}_{-0.9}$ & 
$1.9^{+0.9}_{-0.9}$ \\

$\sigma_{HIRES}$ (\ms) & 
$7.9^{+0.7}_{-0.6}$ & 
$7.8^{+0.7}_{-0.6}$ & 
$7.9^{+0.7}_{-0.6}$ & 
$7.8^{+0.6}_{-0.6}$ \\

$\gamma_{APF}$ (\ms) & 
$-2.7^{+1.0}_{-1.0}$ & 
$-2.7^{+1.0}_{-1.0}$ & 
$-2.7^{+1.0}_{-1.0}$ & 
$-2.7^{+1.0}_{-1.0}$ \\

$\sigma_{APF}$ (\ms) &
$7.3^{+0.5}_{-0.5}$ & 
$7.3^{+0.5}_{-0.5}$ & 
$7.3^{+0.5}_{-0.5}$ & 
$7.3^{+0.5}_{-0.5}$ \\
\hline
\end{tabular}
\end{table*}

\section{Discussion} \label{sec:discussion}

In this section, we discuss the impact of our joint RV-direct imaging analysis on the probable age of the system and the possible planet-disk interactions. We also discuss the prospect of detecting additional planets with future facilities such as the James Webb Space Telescope.

\subsection{Choice of evolutionary models}\label{sub:models}

\begin{figure}
    \centering
    \includegraphics[width=0.5\textwidth]{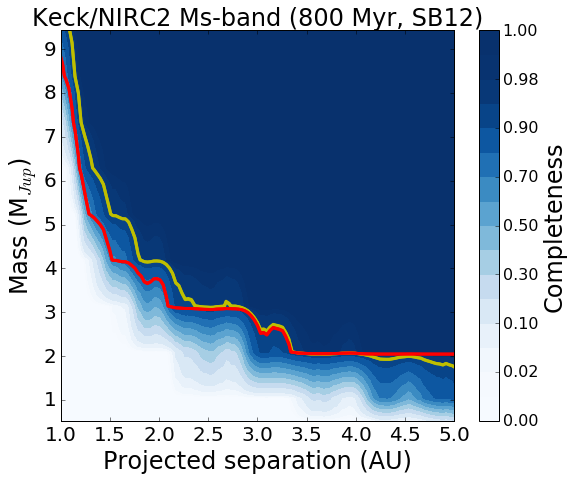}
    \caption{Keck/NIRC2 Ms-band vortex performance/completeness map for a $\tau=5\sigma$ detection threshold comparing using the SB12 evolutionary model \citep{Spiegel2012}. The yellow curve highlights the 95\% completeness contour. The red curve highlights the 95\% completeness contour for the COND model as in Figure~\ref{fig:sensitivity_nirc2}.}
    \label{fig:sb12}
\end{figure}

The direct imaging upper limits are model dependent. We chose to use the COND model mostly for practical reasons. This choice was also motivated by the fact that at the system's age and probable planet mass, evolutionary models have mostly forgotten initial conditions so that hot and cold start models have converged \citep{Marley2007}. However, COND is arguably one of the oldest evolutionary models available. The treatment of opacities, chemistry, etc., are all somewhat outdated. For our 800 Myr case, the most probable age for the system, we generated completeness maps using the evolutionary model presented in \citet{Spiegel2012} as well, SB12 hereafter (Figure~\ref{fig:sb12}). Because the publicly available SB12 grid does not fully cover our age and mass range, some minor extrapolations were necessary. The result of this comparison shows some noticeable discrepancies across the range probed by our data (see Figure~\ref{fig:sb12}). However, both models seem to agree to within error bars at the location of the planet around 3.48 AU, so the impact of the choice of evolutionary model on our joint statistical analysis is only marginal.

\subsection{Constraints on the system's age and inclination}

The planet is not detected in our deep imaging data to the $5\sigma$ threshold. According to our upper limits and RV results, the imaging nondetection indicates that the true age of \thisstar\ is likely to be closer to 800 Myr. Moreover, spectroscopic indicators of age (\logrhk\ and rotation) point towards this star being at the older end of the age range tested here, nearer to 800 Myr than 200 Myr \citep{Mamajek2008}. VLTI observations were used to interferometrically measure the stellar radius and place the star on isochrone tracks. These models also yield an age of 800 Myr or more \citep{Difolco2004}. Thus, the most likely model included here is the 800 Myr model, which is also a fortiori the least restrictive in placing an upper limit on the planet mass. This model yields a mass estimate of $M_{b} = 0.78 ^{+0.38}_{-0.12}\ M_{Jup}$, and an orbital plane inclination of $i = 89^{\circ}\pm 42^{\circ}$.

We note that this inclination is marginally consistent with being co-planar with the outer disk belt, which has a measured inclination of $i=34^{\circ}\pm2^{\circ}$ \citep{Booth2017}. Although the direct imaging nondetection naturally favors near edge-on solutions, the full posterior distribution can still be interpreted as consistent with the planet being co-planar with the outer disk.
With the joint RV and imaging analysis, we are unable to definitively state whether the planet is or is not co-planar with the outer debris disk. However, we are able to rule out ages at or below 200 Myr if coplanarity is required.

To assess the properties of the planet assuming coplanarity with the disk, we repeat our joint analysis implementing a new Gaussian prior on the inclination of $i=34^{\circ}\pm2^{\circ}$, rather than the uninformative geometric prior used in the previous analysis. We ran this analysis for all of the age models. Each of the models satisfied the convergence criterion for all parameters except for longitude of the ascending node ($\Omega$), which is poorly constrained by our data in all cases. However, the 200 Myr model failed to converge in many other parameters, most likely due to the conflict between the youngest models and the coplanarity condition. We therefore report only the 400 Myr and older model results here.

The majority of the fit parameters have posterior distributions consistent with the previous analyses. Since inclination does not correlate strongly with any parameters except companion mass, we do not expect this new prior to affect the posterior probability distributions for any of the other model parameters. 

\begin{figure}
    \centering
    \includegraphics[width=0.5\textwidth]{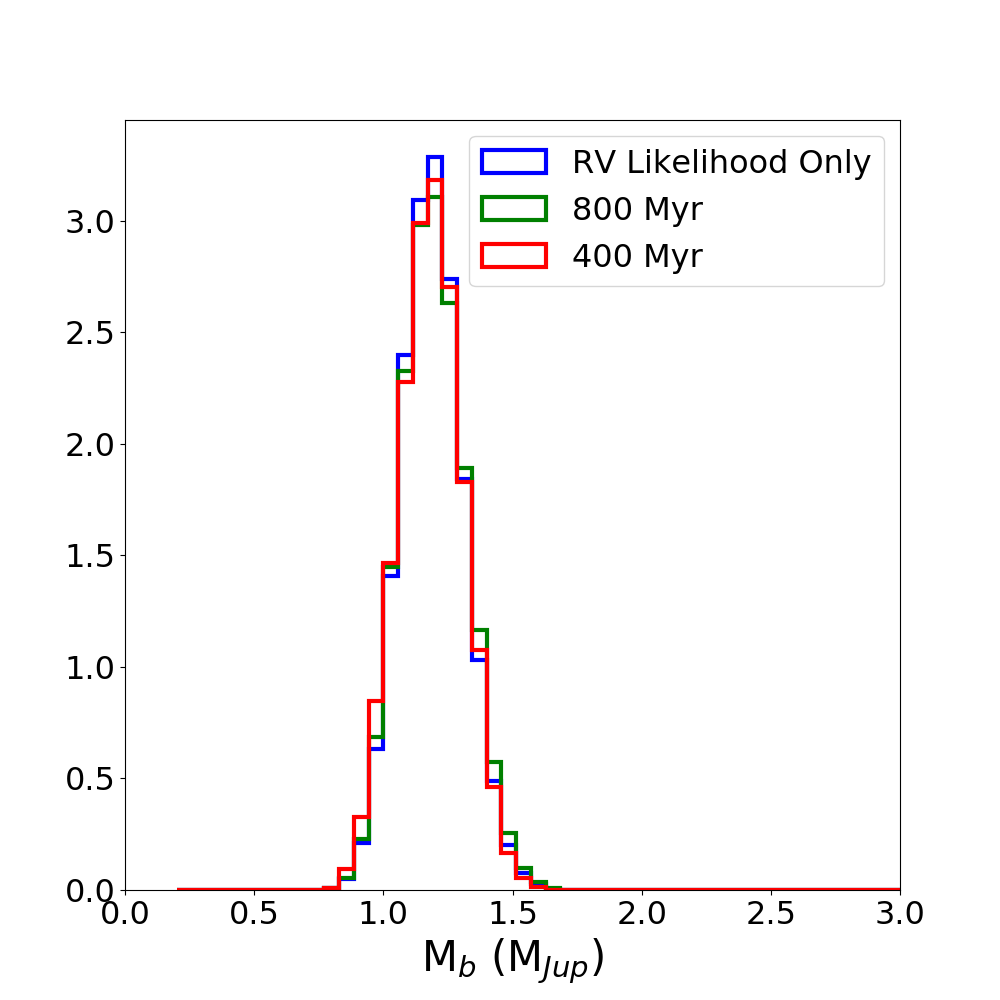}
    \caption{Posterior probability distributions for the mass of \thisstar\ b, when a Gaussian prior preferring orbits coplanar with the measured inclination of the disk ($i = 34^{\circ} \pm 2^{\circ}$) is applied instead of the uninformative geometric prior typically used. The resultant distributions are normalized in the plot. The posterior probability distributions are mutually consistent with one another, since $M_{b}\sin{i}$ is well constrained by the RV data. If coplanarity is assumed, the mass of the planet is $1.19\pm0.12$ \mjup.}
    \label{fig:i34_m2dist}
\end{figure}

For all age models, the companion mass constraint changes significantly compared to the edge-on orbits preferred by the uninformative prior. The new constraints are all similar to one another, with the exception of the 200 Myr model, which did not achieve convergence. The new median and 68\% credible intervals on the planet mass for each model are $m_{b} = 1.19\pm0.12 \mjup$ (RV only); $1.18^{+0.12}_{-0.11} \mjup$ (800 Myr); and $1.19^{+0.11}_{-0.12} \mjup$ (400 Myr). The posterior distributions  are plotted in Figure \ref{fig:i34_m2dist}.

\subsection{Planet-disk interactions}

In this section, we investigate the possible connection between \thisstar\ b and the system's debris belts. Debris disks are the leftovers of planet formation and primarily made out of small bodies (asteroids, comets, dust) that are the remnants of planetesimals, the building blocks of planet cores. Debris disks can be traced by the intrinsic thermal emission of micron-sized particles generated in collisional cascades between the small bodies. For some favorable geometries, scattering off of the small particles can also be detected at optical wavelengths. The dust is constantly replenished by the collisions between leftover planetesimals that are gravitationally stirred by themselves or by neighboring planets. Non-axisymmetric features and separated rings revealed in scattered light images and/or thermal emissions of micron-sized to mm-sized dust particles are considered signposts of existing planets. The connection between debris disks and planets has been seen in several of the currently known directly imaged planetary systems: e.g., HR 8799, $\beta$ Pictoris, HD 95086, HD 106906, Fomalhaut, 51 Eridani (see, e.g., \citet{Bowler2016} for a recent review). Using the largest sample of debris disks systems directly surveyed for long-period giant planets to date, \citet{Meshkat2017} recently found that the occurrence rate of long-period giant planets in dusty systems is about ten times higher than in dust-free systems (88\% confidence level), providing the first tentative empirical evidence for a planet-disk connection. 

\subsubsection{Constraining the inner belt(s)}

\begin{figure}
    \centering
    \includegraphics[width=0.45\textwidth]{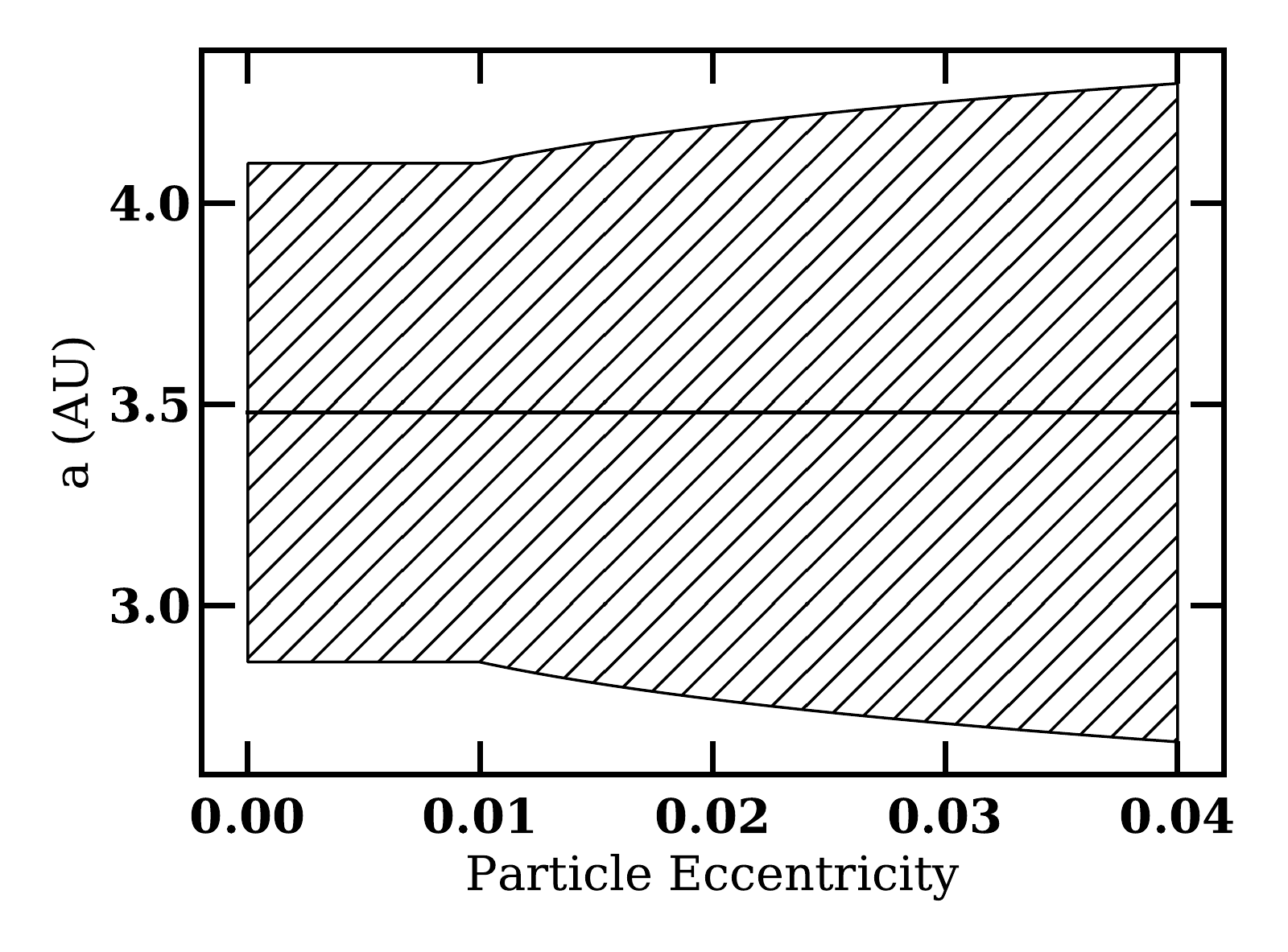}
    \caption{The chaotic zone around \thisstar\ b as a function of particle free eccentricity. The hatched zone corresponds to the chaotic zone around the most likely location of the planet (a = 3.48 AU; horizontal solid line). The planet mass is fixed to 0.78 $M_{\rm Jup}$. We expect the region between $\sim$2.7 and $\sim$4.3 AU to be cleared out by \thisstar\ b.}
    \label{fig:chaotic_zone}
\end{figure}

Armed with refined orbital parameters of \thisstar\ b, we place constraints on the inner belt(s) identified by \citet{Su2017}. Close to the planet, inside a chaotic zone, an overlap of mean motion resonances destabilizes the orbits of particles on short timescales, effectively clearing them out \citep[e.g.,][]{Wisdom80, Quillen06, Morrison15}. \citet{Wisdom80} showed that the width of the chaotic zone scales with $\mu^{2/7}$ where $\mu$ is the mass ratio between the planet and the star. \citet{Quillen06} find that the width of the chaotic zone is independent of planet eccentricity $e_{\rm pl}$ as long as $e_{\rm pl} \lesssim 0.3$. We have shown that \thisstar\ b has an eccentricity consistent with zero, and definitively smaller than 0.3. Using the numerically calibrated expressions for the chaotic zone's inner edge $(1-1.17\mu^{0.28})a_{\rm pl}$ and the outer edge $(1+1.76\mu^{0.31})a_{\rm pl}$ where $a_{\rm pl}$ is the semimajor axis of the planet \citep[see][their Table 1]{Morrison15}, we find that \thisstar\ b would clear out 2.90--4.19 AU if the planet was located at 3.48 AU (equivalent to 0.78 $M_{\rm Jup}$ planet orbiting a star of 0.78 $M_\odot$ at orbital periods of 7.37 yrs; see ``800 Myr'' column of Table \ref{tab:mcmc_res}).

If the particles within the planetesimal belts have sufficiently large free eccentricities (i.e., initial eccentricities set by self-stirring and collision; see, e.g., \citealt{Pan12}), the chaotic zone can widen, commensurate with the particle eccentricities \citep{Mustill12}. For \thisstar\ b, the critical particle eccentricity is only $\sim$ $0.21\mu^{3/7} \sim 0.01$. In Figure~\ref{fig:chaotic_zone}, we show that for particles more eccentric than $\sim$0.01, the chaotic zone around \thisstar\ b can widen by a few tens of percent.

Overall, we expect there to be no particles between $\sim$2.7 and $\sim$4.3 AU. If the excess IR emission interior to 25 AU is from one broad disk, its inner edge must be beyond $\sim$4.3 AU. If instead the excess emission is from two narrow belts, the innermost belt must be inside $\sim$2.7 AU while the outermost belt must be outside $\sim$4.3 AU. Both scenarios are roughly consistent with the analysis of \citet{Su2017} and the most recent LBTI results \citep{Ertel2018}.

\subsubsection{Dust production in planetesimal belts:\\ planet-stirred or self-stirred?}

\begin{figure}
    \centering
    \includegraphics[width=0.45\textwidth]{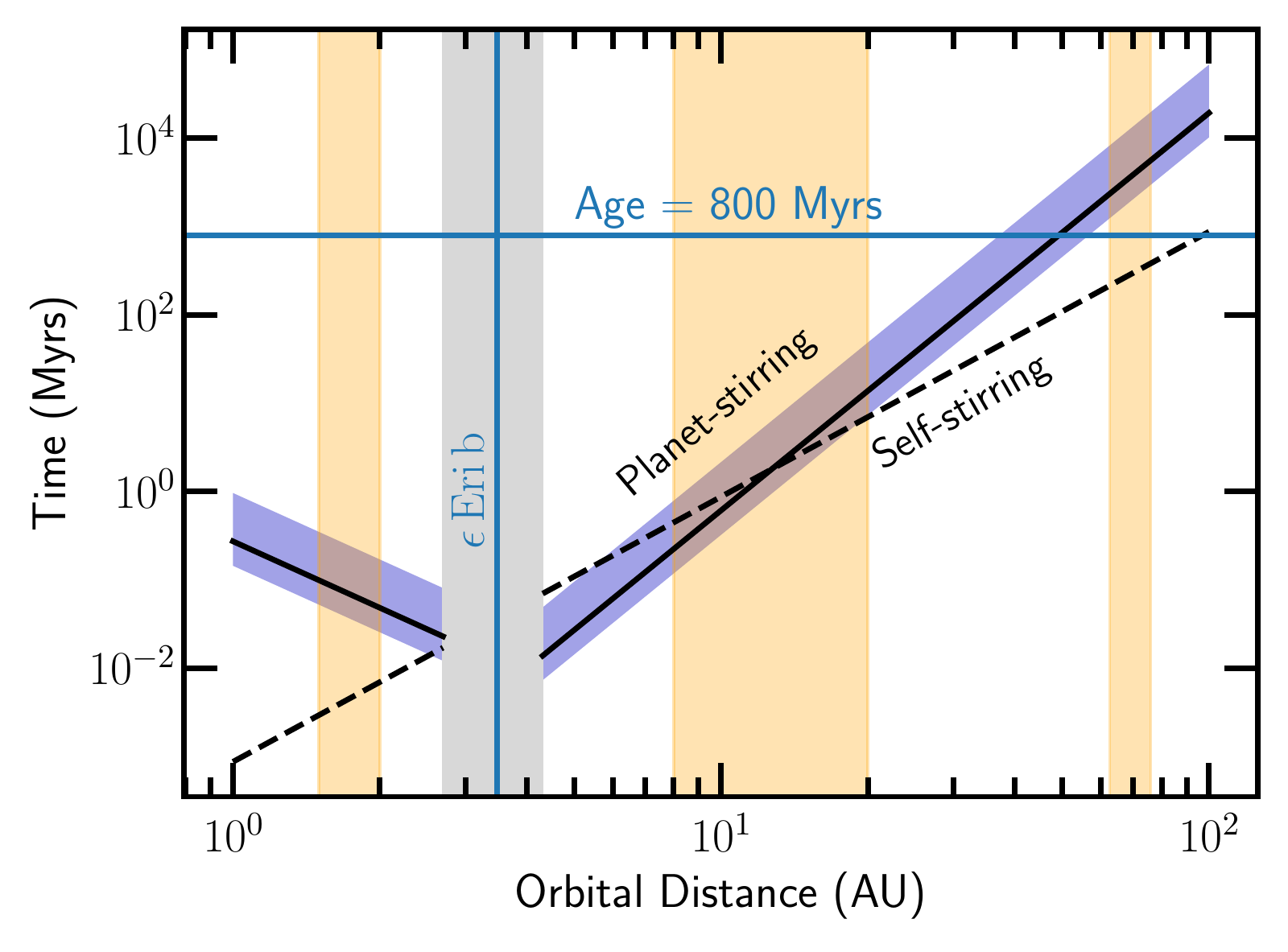}
    \caption{The timescale for \thisstar\ b to stir particles at different distances (equations \ref{eq_cross_int} and \ref{eq_cross_ext}) compared against self-stirring timescale (equation \ref{eq:t_self_stir}). The blue bar reflects the range of eccentricities of \thisstar\ b from our MCMC analysis, with the lower and upper limit corresponding to $e_{\rm pl} = 0.13$ and $e_{\rm pl} = 0.02$, respectively; the solid line represents the median $e_{\rm pl} = 0.07$. We fix $M_{\rm pl}=0.78\,M_{\rm Jup}$, $a_{\rm pl}=3.48\,{\rm AU}$, and $M_\star=0.78\,M_\odot$. We have assumed minimum mass Solar nebular (MMSN) surface density to calculate the self-stirring timescale. The expected chaotic zone of \thisstar\ b is depicted with a gray zone. Three orange bars illustrate the warm and cold belts reported by \citet{Su2017} and \citet{Booth2017}, respectively. Unless \thisstar's disk is significantly less massive than MMSN---by more than an order of magnitude---it is likely that the collisions within the innermost warm belt and the outermost cold belt are driven by self-stirring rather than the secular perturbation by \thisstar\ b. Collisions in the secondary warm belt that span 8--20 AU are consistent with both self-stirring and planet-stirring.}
    \label{fig:stirring}
\end{figure}

Dust grains that are mm-sized and smaller that dominate the radio and scattered light images are the products of collisions among larger planetesimals. Particles can collide with each other as they gravitationally stir each other or they may be secularly perturbed by a planet. Is \thisstar\ b responsible for triggering collisional cascades in the warm and cold belts detected in the system or are the planetesimals more likely to be self-stirred?

The characteristic timescale for planetesimals to stir each other is set by their rate of collision (see e.g., Section 4.1.4 of \citealt{Goldreich04}). Numerical \citep[e.g.,][]{Kenyon08} and analytic \citep[e.g.,][]{Pan12} calculations suggest bodies as large as $\sim$1000 km are at the top of the collisional cascade: they can stir up smaller bodies to disruption at a rate faster than their growth via accretion.
It is the formation timescale of these Pluto-sized bodies that limits the overall self-stirring timescale:
\begin{equation}
\label{eq:t_self_stir}
    t_{\rm ss} \sim 173\,{\rm Myrs}\left(\frac{\Sigma_{\rm MMSN}}{\Sigma}\right)^{1.15}\left(\frac{a_{\rm disk}}{60\,{\rm AU}}\right)^3\left(\frac{M_\odot}{M_\star}\right)^{3/2},
\end{equation}
where $a_{\rm disk}$ is the semimajor axis of the planetesimals, $M_\star$ is the mass of the host star, $\Sigma_{\rm MMSN} \equiv 0.18\,{\rm g\,cm^{-2}}(M_\star/M_\odot)(a/30\,{\rm AU})^{-3/2}$ is the minimum mass Solar nebula, and we take the coefficient derived by numerical simulations of \citet{Kenyon08}. 
The calculation above assumes large bodies build up mostly after the dispersal of the gas disk since gas dynamical friction tend to damp away the eccentricities of solids. It may be that planetesimals are built rapidly even in the early stages of gas disk evolution---by streaming instability \citep[e.g.,][]{Youdin05,Johansen07}, resonant drag instability \citep{Squire18}, and/or pebble accretion \citep[e.g.,][]{Ormel10,Ormel17}---so that Pluto-sized bodies are already available by the time the disk gas dissipates. Without the need to wait for the build-up of large bodies after the gas dispersal, $t_{\rm ss}$ can be shortened by an order of magnitude \citep{Krivov18}.

The orbit crossing timescale for two nearby planetesimals as they are secularly perturbed by a planet is $t_{\rm cross} \sim 1/Ae_{\rm p}$ where $A$ is the precession frequency and $e_{\rm p}$ is the eccentricity of the perturbed particle \citep{Mustill2009}. For particles on initially circular orbits, $e_{\rm p}$ is set by the forced eccentricity so that the orbit-crossing timescale becomes
\begin{eqnarray}\label{eq_cross_int}
 t_{\rm cross} \approx 500\,{\rm Gyrs}\,\frac{(1-e_{\rm pl}^2)^{3/2}}{e_{\rm pl}} \left(\frac{a_{\rm disk}}{60\ AU} \right)^{9/2} \nonumber \\
 \times \left(\frac{M_\star}{M_\odot} \right)^{1/2} 
  \left(\frac{M_{\rm Jup}}{M_{\rm pl}} \right) 
  \left(\frac{3.5\,{\rm AU}}{a_{\rm pl}} \right)^{3}
 \end{eqnarray}
for internal perturbers, and
\begin{eqnarray}\label{eq_cross_ext}
 t_{\rm cross} \approx 17.4\,{\rm kyrs}\,\frac{(1-e_{\rm pl}^2)^{3/2}}{e_{\rm pl}} \left(\frac{a_{\rm pl}}{3.5\,{\rm AU}} \right)^4 \nonumber \\
 \times \left(\frac{M_\star}{M_\odot} \right)^{1/2} 
  \left(\frac{M_{\rm Jup}}{M_{\rm pl}} \right)
 \left(\frac{1\,{\rm AU}}{a_{\rm disk}} \right)^{5/2}
 \end{eqnarray}
for external perturbers, where $e_{\rm pl}$ is the planet eccentricity, $M_{\rm pl}$ is the planet mass, and $a_{\rm pl}$ is the semimajor axis of the perturbing planet \citep[see][their equations 15 and 16]{Mustill2009}. For \thisstar\ b to be responsible for the dust production in planetesimal belts, $t_{\rm cross}$ must be shorter than the age of the system and $t_{\rm cross} < t_{\rm ss}$. For our refined parameters for \thisstar\ b, we find that out of the three belts that are imaged or inferred to exist, the innermost and the outermost belts are more likely self-stirred while the intermediate warm belt can be either self-stirred or planet-stirred (see Figure~\ref{fig:stirring}). 

Even the self-stirring timescale for the outer belt is uncomfortably close to the system's estimated age. It may be that the largest bodies that trigger a collisional cascade emerge rapidly such that our $t_{\rm ss}$ is overestimated. It may also be that an extra planet beyond the orbit of \thisstar\ b is responsible for stirring the outermost belt and for carving out the region between the inner warm belts and the outer cold belt. 
Our solution with uninformative inclination priors to the orbit of \thisstar\ b favors an orbital inclination ($89^\circ \pm 42^\circ$) that is larger than the inclination of the outermost belt ($34^\circ \pm 2^\circ$).
Such a large misalignment suggests that \thisstar\, may have interacted with a fly-by in the past and/or there exists another planet that is inclined to the orbit of \thisstar\ b.

Should this extra planet be responsible for tilting the outer belt from the orbital plane of \thisstar\, b, we expect its orbit to be misaligned with respect to \thisstar\,b by at least 10$^\circ$ assuming the mass of the extra planet is $\sim$1 Jupiter mass with orbital distance of 48 AU.

In \citet{Meshkat2017}, we found that the occurrence rate of long-period planets in debris disks is about 10 times higher than in dust-free systems. The \thisstar\, system is rich in dust contained in multiple belts reminiscent of the HR 8799 system \citep{Su2009}, so it may very well harbor more than one planet,
as suggested by previous studies \citep{Mizuki2016,Booth2017}. Unfortunately, the field of view of our high contrast Keck/NIRC2 dataset is too small to probe large separation in the vicinity of the outer disk. 

\subsubsection{Imaging the elusive \thisstar\, planet(s) with JWST}

The Keck/NIRC2 Ms-band vortex coronagraph high contrast images presented here showcase exquisite inner working angle ($0\farcs 2$) and sensitivity up to $5\arcsec$. Ground-based adaptive optics and small angle coronagraphy in the mid-infrared (from 3 to 5 microns) on 10-meter class telescopes will only be challenged by the advent of Giant Segmented Mirror Telescopes. However, NIRCam and MIRI, the infrared cameras of the upcoming James Webb Space Telescope will have unmatched sensitivity beyond $1\arcsec$ from the star. Thus, JWST's NIRCam and MIRI are ideal instruments to explore the inner cavity of \thisstar\ from 3 to 30 microns, and reveal additional elusive planets. The outer ring being at $20\arcsec$ from the host star, the diffraction and scattering from the star will be less of a concern, allowing us to probe the cavity for the putative planet shepherding the outer ring. Invoking dynamical arguments similar to those used in the previous sections and upper limits from Spitzer images \citep{Janson2015}, \citet{Booth2017} predicts that the putative exoplanet responsible for shaping the outer belt is at a semi-major axis of 48 AU, with a mass between 0.4 and 1.2 $M_{J}$.

JWST's NIRCam instrument team is planning coronagraphic observations at 4.4 $\mu$m (F444W) to search both the interior ($\pm$10\arcsec)  and exterior regions (2.2\arcmin) around \thisstar\ for planets. The interior region will be observed twice to reject background objects on the basis of source motion while the entire field will be observed at 3 $\mu$m to reject background stars and galaxies on the basis of color. The expected contrast ratio after roll and reference star subtraction is expected to be $\sim10^{-6}$ at 1\arcsec\ and $\sim10^{-7}$ at 2\arcsec\ depending on the in-orbit stability of JWST \citep{Krist2007,Beichman2010}.  

As noted above, translating between instrumental detection limits and planet mass is challenging due to the uncertainties in stellar age and exoplanet models at low masses. For a nominal age of 800 Myr, SB12 and COND models yield F444W brightness estimates ranging from 16 to 18 mag for a 0.78 M$_{Jup}$ planet (effective temperature of $\simeq 150$ K). In the most favorable case (SB12), NIRCam could detect \thisstar\ b at 1\arcsec\ (close to the nominal 7.37 year, 3.48 AU orbit) with signal-to-noise SNR $\sim$5 and beyond 2\arcsec\ with SNR $>$ 25. Within 1\arcsec\ the Keck observations reported here are comparable to or more sensitive than planned JWST observations due to Keck's larger aperture and the improved performance of the vortex coronagraph. At larger separations, however, the low thermal background in space gives JWST a significant advantage to look for $<$ 0.5 M$_{Jup}$ planets (depending on model assumptions) which might be responsible for the disk structures seen in the far-infrared and by ALMA. Since the field of view of both NIRCam and MIRI is $20\arcsec \times 20\arcsec$, at least two pointings will be necessary to map out the inner cavity at the vicinity of the outer ring.

\section{Conclusion} \label{sec:conclusions}

In this paper, we presented the most sensitive and comprehensive observational evidence for the existence of \thisstar\ b. Combining exquisite RV and direct imaging data provides unprecedented constraints on the mass and orbital parameters of the planet. \thisstar\ b has a mass of $0.78^{+0.38}_{-0.12}$ $M_{Jup}$ and is orbiting \thisstar\ at about $3.48\pm 0.02$ AU in $7.37 \pm 0.07$ years. Assuming coplanarity with the outer belt resolved by ALMA (inclination of $34^\circ \pm 2^\circ$), the mass of the planet is approximately 1.19 $M_{Jup}$. Our data and analysis also presents compelling lines of evidence that the system's age is closer to 800 Myr or more. Indeed, the direct imaging data should have shown the planet if the system's age was 200 Myr. Notably, we also find that the eccentricity of \thisstar\ b is a very low $0.07^{+0.06}_{-0.05}$, an order of magnitude smaller than early estimates and consistent with a circular orbit. For that reason and using simple dynamical arguments, we postulate that \thisstar\ b is unlikely to be responsible for stirring the outer debris belt at 70 AU, and that one or more additional planets must be shepherding it. However, \thisstar\ b could be responsible for shaping the putative warm belt(s) within 25 AU, although self-stirring is another likely dust production mechanism. If the additional planet(s) exist(s), they will be easily detected with JWST's NIRCam and MIRI from 4 to 25 microns, enabling unique spectroscopic and dynamical characterization opportunities.

This paper also demonstrates the unique power of the combination of radial velocity and direct imaging observations to detect and constrain the masses of giant planets within 10 AU. The long history of RV observations limited the spatial domain in which \thisstar\ b could reside, allowing optimized detectability with the imaging observations.

We would be remiss if we did not mention as a caveat that the direct imaging nondetection of the planet leaves open the possibility, however small, that the planet does not actually exist. Although there is no obvious commensurability between the long-term magnetic activity diagnostic (\shk) and the RV data, as demonstrated in \sectionautorefname \ref{subsec:robustness}, it remains possible that the magnetic field affects the RVs in a way we don't understand. 

\thisstar\ remains a fascinating testbed for studying planetary formation in great detail. We conclude this paper by emphasizing the perfect complementarity between long-term radial velocity monitoring, mid-infrared small-angle high-contrast ground-based capabilities, and the sensitive space-based parameter space to be opened by JWST.

\acknowledgments 

The authors would like to acknowledge the contributions and useful comments from Prof.~Debra Fischer and Prof.~James Graham. The data presented herein were obtained at the W.M. Keck Observatory, which is operated as a scientific partnership among the California Institute of Technology, the University of California and the National Aeronautics and Space Administration (NASA). The Observatory was made possible by the generous financial support of the W.M. Keck Foundation. The authors wish to recognize and acknowledge the very significant cultural role and reverence that the summit of Maunakea has always had within the indigenous Hawaiian community.  We are most fortunate to have the opportunity to conduct observations from this mountain.  This work was partially performed at the Jet Propulsion Laboratory, California Institute of Technology, under contract the the National Aeronautics and Space Administration. (C) 2017. All rights reserved. E.C. acknowledges support from NASA through Hubble Fellowship grant HF2-51355 awarded by STScI, which is operated by AURA, Inc. for NASA under contract NAS5- 26555. We also acknowledge support from NASA/NExSS  through grant number NNX15AD95G.

\software{\radvel \citep{Fulton2018}, \emcee \citep{Foreman-Mackey2013}, \VIP \citep{Gomez2017}, \PyKLIP \citep{Wang2015}}

\clearpage

\begin{ThreePartTable}
\begin{longtable}{cccc}
\caption{Keck Radial Velocity Measurements \label{tab:rvs_k}}\\
\hline
\textbf{JD} & \textbf{RV (m/s)\tnote{$^a$}} & \textbf{$\sigma_{RV}$ (m/s)\tnote{$^b$}} & \textbf{\shk} \\
\hline
\endfirsthead
\multicolumn{4}{c}%
{\tablename\ \thetable\ -- \textit{Continued from previous page}} \\
\hline
\textbf{JD} & \textbf{RV (m/s)\tnote{$^a$}} & \textbf{$\sigma_{RV}$ (m/s)\tnote{$^b$}} & \textbf{\shk} \\
\hline
\endhead
\hline \multicolumn{4}{r}{\textit{Continued on next page}} \\
\endfoot
\hline
\endlastfoot
2455110.97985 & -6.54 & 1.30 & 0.467 \\
2455171.90825 & -3.33 & 1.09 & 0.486 \\
2455188.78841 & 7.90 & 1.11 & 0.481 \\
2455231.7593 & -8.39 & 1.13 & 0.497 \\
2455255.70841 & 1.66 & 0.70 & 0.520 \\
2455260.71231 & 1.77 & 1.01 & 0.523 \\
2455261.71825 & 0.75 & 1.30 & 0.526 \\
2455413.14376 & -10.67 & 0.76 & 0.500 \\
2455414.13849 & -16.73 & 0.99 & 0.000 \\
2455415.14082 & -20.89 & 0.78 & 0.495 \\
2455426.14477 & -17.57 & 0.86 & 0.494 \\
2455427.14813 & -18.05 & 0.87 & 0.483 \\
2455428.14758 & -21.46 & 0.87 & 0.480 \\
2455429.14896 & -18.67 & 0.90 & 0.475 \\
2455434.14805 & 7.21 & 0.86 & 0.474 \\
2455435.14705 & 4.46 & 0.89 & 0.481 \\
2455436.14535 & -2.48 & 0.83 & 0.485 \\
2455437.15006 & -5.03 & 0.94 & 0.480 \\
2455438.15172 & -14.24 & 0.90 & 0.484 \\
2455439.14979 & -13.17 & 0.51 & 0.474 \\
2455440.15188 & -22.38 & 0.88 & 0.471 \\
2455441.15033 & -19.71 & 0.99 & 0.469 \\
2455456.01632 & 4.52 & 0.97 & 0.466 \\
2455465.07401 & -12.99 & 0.98 & 0.449 \\
2455469.1284 & 7.81 & 1.01 & 0.465 \\
2455471.97444 & -4.15 & 1.16 & 0.471 \\
2455487.00413 & -9.44 & 0.96 & 0.454 \\
2455500.98687 & -2.23 & 1.05 & 0.461 \\
2455521.89317 & -11.42 & 1.05 & 0.455 \\
2455542.95125 & -8.56 & 1.20 & 0.458 \\
2455613.70363 & 0.65 & 1.01 & 0.466 \\
2455791.13884 & 1.87 & 0.87 & 0.433 \\
2455792.13464 & -9.19 & 0.90 & 0.430 \\
2455793.13858 & -17.85 & 0.89 & 0.426 \\
2455795.14053 & -15.43 & 0.96 & 0.418 \\
2455797.13828 & -5.67 & 0.83 & 0.419 \\
2455798.14195 & -5.00 & 0.84 & 0.424 \\
2455807.1116 & -3.91 & 0.99 & 0.417 \\
2455809.1367 & -0.90 & 0.99 & 0.429 \\
2455870.9902 & 1.81 & 1.20 & 0.437 \\
2455902.82961 & 4.20 & 0.74 & 0.429 \\
2455960.69933 & -8.22 & 1.21 & 0.460 \\
2456138.12976 & -2.69 & 0.86 & 0.464 \\
2456149.05961 & -2.49 & 0.53 & 0.470 \\
2456173.13157 & -1.22 & 0.96 & 0.459 \\
2456202.99824 & 19.64 & 0.71 & 0.507 \\
2456327.70174 & 20.33 & 1.05 & 0.535 \\
2456343.7026 & 16.52 & 1.05 & 0.505 \\
2456530.11763 & 6.76 & 0.90 & 0.489 \\
2456532.12218 & 8.06 & 0.85 & 0.479 \\
2456587.96668 & 14.41 & 1.03 & 0.479 \\
2456613.91026 & 15.04 & 1.02 & 0.481 \\
2456637.81493 & 23.88 & 1.02 & 0.487 \\
2456638.79118 & 32.35 & 1.07 & 0.491 \\
2456674.80603 & 11.70 & 1.03 & 0.488 \\
2456708.78257 & 2.49 & 0.99 & 0.482 \\
2456884.13093 & 12.85 & 0.95 & 0.446 \\
2456889.14678 & 18.51 & 0.82 & 0.466 \\
2456890.14703 & 13.09 & 0.86 & 0.461 \\
2456894.13998 & 8.71 & 0.83 & 0.446 \\
2456896.11131 & 15.09 & 0.78 & 0.447 \\
2456910.94964 & 13.84 & 0.64 & 0.450 \\
2457234.13834 & 9.97 & 0.85 & 0.491 \\
2457240.99109 & 6.26 & 0.52 & 0.468 \\
2457243.14297 & 3.19 & 0.78 & 0.476 \\
2457245.14532 & 5.26 & 0.90 & 0.479 \\
2457246.14242 & -1.45 & 0.99 & 0.477 \\
2457247.14678 & -5.60 & 1.01 & 0.482 \\
2457254.14889 & 8.50 & 0.80 & 0.475 \\
2457255.15244 & 6.36 & 0.91 & 0.466 \\
2457256.15168 & 5.80 & 0.83 & 0.476 \\
2457265.14924 & 5.74 & 0.88 & 0.469 \\
2457291.04683 & 6.07 & 1.05 & 0.491 \\
2457326.9831 & 6.10 & 1.12 & 0.501 \\
2457353.88153 & -0.55 & 1.09 & 0.519 \\
2457378.78993 & 2.19 & 1.08 & 0.519 \\
2457384.78144 & 14.17 & 1.10 & 0.517 \\
2457401.75106 & 6.07 & 0.99 & 0.517 \\
2457669.02614 & 1.91 & 1.10 & 0.497 \\
2457672.99494 & -1.33 & 1.20 & 0.497 \\
2457678.97973 & -13.88 & 1.10 & 0.495 \\
2457704.03411 & -14.12 & 0.67 & 0.501 \\
2457712.99284 & -4.84 & 1.18 & 0.478 \\
2457789.74988 & -13.12 & 1.12 & 0.439 \\
2457790.737 & -8.09 & 1.01 & 0.440 \\
2457803.70407 & -4.25 & 1.09 & 0.460 \\
2457804.70718 & -6.55 & 1.09 & 0.471 \\
2457806.79201 & -11.62 & 1.13 & 0.464 \\
2457828.7545 & -12.69 & 1.12 & 0.455 \\
2457829.71875 & -19.82 & 0.98 & 0.466 \\
2457830.71979 & -12.66 & 1.10 & 0.465 \\
\end{longtable}
\begin{tablenotes}
\item[$^a$]{The RV data points listed in these tables are not offset-subtracted.}
\item[$^b$]{Uncertainties quoted in these tables reflect the internal statistical variance of the spectral chunks used to extract the RV data points (see \S \ref{subsec:doppler_analysis} for a full description). They do not include jitter.}
\end{tablenotes}
\end{ThreePartTable}

\clearpage

\begin{ThreePartTable}
\begin{longtable}{cccc}
\caption{APF Radial Velocity Measurements \label{tab:rvs_a}}\\
\hline
\textbf{JD} & \textbf{RV (m/s)\tnote{$^a$}} & \textbf{$\sigma_{RV}$ (m/s)\tnote{$^b$}} & \textbf{\shk} \\
\hline
\endfirsthead
\multicolumn{4}{c}%
{\tablename\ \thetable\ -- \textit{Continued from previous page}} \\
\hline
\textbf{JD} & \textbf{RV (m/s)\tnote{$^a$}} & \textbf{$\sigma_{RV}$ (m/s)\tnote{$^b$}} & \textbf{\shk} \\
\hline
\endhead
\hline \multicolumn{4}{r}{\textit{Continued on next page}} \\
\endfoot
\hline
\endlastfoot
2456582.93034 & 26.64 & 2.73 & 0.524 \\
2456597.91368 & 6.40 & 2.36 & 0.528 \\
2456606.68427 & 16.52 & 0.75 & 0.531 \\
2456608.10376 & 4.69 & 0.78 & 0.530 \\
2456610.7625 & 16.04 & 1.18 & 0.512 \\
2456618.88476 & -2.11 & 0.78 & 0.530 \\
2456624.72004 & 4.20 & 1.11 & 0.519 \\
2456626.81421 & 24.46 & 0.75 & 0.521 \\
2456628.72976 & 24.14 & 0.70 & 0.540 \\
2456631.42746 & -2.26 & 0.88 & 0.502 \\
2456632.80921 & 14.46 & 0.62 & 0.523 \\
2456644.75696 & 8.20 & 2.30 & 0.522 \\
2456647.81171 & 14.44 & 0.63 & 0.535 \\
2456648.59184 & 12.62 & 1.10 & 0.538 \\
2456662.63738 & 9.77 & 0.73 & 0.536 \\
2456663.75415 & 10.43 & 1.11 & 0.531 \\
2456667.52792 & 18.00 & 0.78 & 0.535 \\
2456671.68695 & 19.96 & 1.05 & 0.604 \\
2456675.75647 & 7.84 & 1.12 & 0.519 \\
2456679.83732 & 17.70 & 1.05 & 0.529 \\
2456682.56608 & 17.80 & 0.82 & 0.550 \\
2456689.76638 & 26.34 & 0.75 & 0.500 \\
2456875.02028 & 7.12 & 2.18 & 0.501 \\
2456894.88054 & 8.28 & 1.30 & 0.470 \\
2456901.06193 & 9.95 & 1.54 & 0.479 \\
2456909.10279 & -4.71 & 1.21 & 0.476 \\
2456922.07953 & 12.25 & 2.13 & 0.461 \\
2456935.94021 & -2.43 & 1.27 & 0.479 \\
2456937.92403 & -0.55 & 1.35 & 0.468 \\
2456950.03798 & 3.82 & 1.44 & 0.472 \\
2456985.64755 & -1.80 & 2.28 & 0.441 \\
2456988.63095 & 5.93 & 1.29 & 0.478 \\
2456999.76434 & 8.84 & 1.37 & 0.459 \\
2457015.72916 & -2.17 & 1.10 & 0.465 \\
2457026.78021 & -1.44 & 1.34 & 0.464 \\
2457058.45996 & -3.69 & 1.89 & 0.435 \\
2457234.08236 & 7.73 & 1.39 & 0.525 \\
2457245.86234 & -4.19 & 1.41 & 0.519 \\
2457249.93007 & -3.94 & 1.31 & 0.500 \\
2457253.11257 & 5.63 & 1.33 & 0.511 \\
2457257.15719 & -1.02 & 1.15 & 0.506 \\
2457258.94437 & -12.69 & 1.23 & 0.517 \\
2457261.02221 & -2.76 & 1.32 & 0.501 \\
2457262.94505 & -7.81 & 1.36 & 0.496 \\
2457265.95783 & 9.67 & 1.24 & 0.516 \\
2457275.01304 & -1.91 & 1.23 & 0.515 \\
2457283.96368 & 1.88 & 1.29 & 0.507 \\
2457287.02735 & -1.11 & 1.35 & 0.524 \\
2457290.95635 & 3.19 & 1.42 & 0.534 \\
2457305.83659 & -5.63 & 1.23 & 0.515 \\
2457308.90844 & 13.30 & 1.26 & 0.534 \\
2457318.83435 & 8.72 & 1.26 & 0.557 \\
2457321.79157 & 6.64 & 1.36 & 0.540 \\
2457325.84352 & 2.87 & 1.41 & 0.543 \\
2457331.10764 & 9.90 & 1.36 & 0.552 \\
2457332.78237 & 9.64 & 1.25 & 0.558 \\
2457334.82998 & 5.22 & 1.30 & 0.548 \\
2457337.7891 & 5.41 & 1.59 & 0.545 \\
2457340.95644 & -1.99 & 1.27 & 0.553 \\
2457347.86896 & 4.10 & 1.29 & 0.556 \\
2457348.77993 & 4.65 & 1.27 & 0.556 \\
2457350.72611 & 5.83 & 1.20 & 0.558 \\
2457354.70613 & -0.88 & 1.65 & 0.548 \\
2457361.64656 & 17.26 & 1.43 & 0.549 \\
2457364.77113 & -7.80 & 1.30 & 0.531 \\
2457365.70544 & 0.72 & 1.26 & 0.550 \\
2457424.71436 & -1.68 & 1.37 & 0.555 \\
2457426.63205 & 3.62 & 1.42 & 0.559 \\
2457427.38923 & 3.97 & 1.17 & 0.577 \\
2457429.72793 & 2.42 & 0.90 & 0.560 \\
2457432.60322 & 6.20 & 1.25 & 0.569 \\
2457435.69406 & -18.61 & 18.79 & 0.304 \\
2457443.66061 & 2.25 & 1.24 & 0.559 \\
2457446.70278 & 3.96 & 1.37 & 0.566 \\
2457471.55712 & 5.85 & 1.63 & 0.535 \\
2457599.93545 & -5.69 & 0.85 & 0.505 \\
2457605.99828 & -5.33 & 1.27 & 0.559 \\
2457607.92844 & -24.97 & 1.39 & 0.540 \\
2457611.16197 & -16.02 & 1.26 & 0.510 \\
2457613.86777 & 2.47 & 1.54 & 0.560 \\
2457615.04307 & 3.50 & 1.48 & 0.538 \\
2457617.08138 & 0.91 & 1.29 & 0.555 \\
2457619.05397 & -12.30 & 1.46 & 0.529 \\
2457621.79772 & -13.43 & 1.57 & 0.508 \\
2457626.10874 & 0.39 & 1.33 & 0.534 \\
2457627.95628 & -4.92 & 1.37 & 0.551 \\
2457633.96762 & -8.24 & 1.70 & 0.512 \\
2457636.08672 & -1.33 & 1.18 & 0.539 \\
2457637.95848 & -7.66 & 1.37 & 0.538 \\
2457643.92459 & -14.39 & 1.33 & 0.512 \\
2457668.93315 & -0.83 & 1.34 & 0.527 \\
2457669.90475 & 2.76 & 1.43 & 0.533 \\
2457670.88203 & -8.82 & 1.42 & 0.543 \\
2457674.61398 & -5.61 & 1.42 & 0.534 \\
2457679.98028 & -12.42 & 1.78 & 0.515 \\
2457687.77138 & 1.17 & 1.37 & 0.524 \\
2457694.76122 & -3.81 & 1.33 & 0.504 \\
2457696.82099 & -5.60 & 1.32 & 0.522 \\
2457700.96748 & -10.84 & 1.41 & 0.534 \\
2457701.84849 & -11.69 & 1.38 & 0.517 \\
2457702.89789 & -14.82 & 1.22 & 0.524 \\
2457703.82658 & -19.89 & 1.25 & 0.523 \\
2457705.73282 & -9.58 & 1.32 & 0.513 \\
2457707.78376 & -9.03 & 1.24 & 0.511 \\
2457717.79818 & -15.06 & 1.22 & 0.505 \\
2457722.75749 & -12.43 & 2.05 & 0.427 \\
2457728.81592 & -7.64 & 1.67 & 0.514 \\
2457741.79955 & -14.52 & 1.16 & 0.513 \\
2457743.5028 & -17.28 & 1.32 & 0.489 \\
2457745.93451 & -17.74 & 1.31 & 0.487 \\
2457749.71344 & -5.63 & 1.30 & 0.503 \\
2457751.64976 & -16.16 & 1.32 & 0.501 \\
2457753.47716 & -12.45 & 1.30 & 0.509 \\
2457798.55461 & -18.91 & 2.25 & 0.465 \\
2457821.65582 & -5.60 & 1.63 & 0.490 \\
\end{longtable}
\begin{tablenotes}
\item[$^a$]{The RV data points listed in these tables are not offset-subtracted.}
\item[$^b$]{Uncertainties quoted in these tables reflect the internal statistical variance of the spectral chunks used to extract the RV data points (see \S \ref{subsec:doppler_analysis} for a full description). They do not include jitter.}
\end{tablenotes}
\end{ThreePartTable}

\clearpage

\acknowledgments
\bibliography{epseri_v2}

\end{document}